\documentclass[sigconf]{acmart}
\usepackage{CJKutf8}
\usepackage[utf8]{inputenc}
\usepackage[T1]{fontenc}
\usepackage{textcomp}
\usepackage{xcolor,colortbl}
\usepackage{multirow}
\usepackage{multicol}
\usepackage{threeparttable}
\usepackage{soul}
\usepackage{hyperref} 
\usepackage{longtable}
\usepackage{subfig}
\usepackage{overpic}
\usepackage{array}

\AtBeginDocument{%
  \providecommand\BibTeX{{%
    \normalfont B\kern-0.5em{\scshape i\kern-0.25em b}\kern-0.8em\TeX}}}

\newif\ifdraft
\drafttrue

\newcommand{\modi}[1]{\textcolor{black}{#1}}

\begin{document}
%\setcopyright{acmcopyright}
%\copyrightyear{2025}
%\acmYear{2025}
%\acmDOI{XX.XXXX/XXXXXXX.XXXXXXX}

%\acmJournal{JACM}
%\acmVolume{1}
%\acmNumber{1}
%\acmArticle{1}
%\acmMonth{1}

\copyrightyear{2025} 
\acmYear{2025} 
\setcopyright{acmlicensed}\acmConference[CHI '25]{Proceedings of the CHI Conference on Human Factors in Computing Systems}{April 26-May 1, 2025}{Yokohama, Japan}
\acmBooktitle{Proceedings of the CHI Conference on Human Factors in Computing Systems (CHI '25), April 26--May 1, 2025, Yokohama, Japan}
\acmDOI{XX.XXXX/XXXXXXX.XXXXXXX}
\acmISBN{}

\title[How Chinese Younger Family Members Safeguard Seniors from Online Fraud]{``Auntie, Please Don't Fall for Those Smooth Talkers'': How Chinese Younger Family Members Safeguard Seniors from Online Fraud}

% \title{``Auntie, Please Don't Fall for Those Smooth Talkers'': How Chinese Younger Family Members Safeguard Seniors from Online Fraud}

% Designing fair hiring pipelineg based on job-seekers' fairness concerns

\author{Yue Deng}
\affiliation{%
  \institution{Hong Kong University of Science and Technology}
  \city{Hong Kong}
  \country{China}
}
\email{ydengbi@connect.ust.hk}

\author{Changyang He}
\affiliation{%
  \institution{Max Planck Institute for Security and Privacy}
  \city{Bochum}
  \country{Germany}
}
\email{changyang.he@mpi-sp.org}

\author{Yixin Zou}
\affiliation{%
  \institution{Max Planck Institute for Security and Privacy}
  \city{Bochum}
  \country{Germany}
}
\email{yixin.zou@mpi-sp.org}

\author{Bo Li}
\affiliation{%
  \institution{Hong Kong University of Science and Technology}
  \city{Hong Kong}
  \country{China}
}
\email{bli@cse.ust.hk}

\renewcommand{\shortauthors}{Deng et al.}

\begin{abstract}
Online fraud substantially harms individuals and seniors are disproportionately targeted. While family is crucial for seniors, little research has empirically examined how they protect seniors against fraud. To address this gap, we employed an inductive thematic analysis of 124 posts and 16,872 comments on RedNote (Xiaohongshu), exploring the family support ecosystem for senior-targeted online fraud in China. We develop a taxonomy of senior-targeted online fraud from a familial perspective, revealing younger members often spot frauds hard for seniors to detect, such as unusual charges. Younger family members fulfill multiple safeguarding roles, including preventative measures, fraud identification, fraud persuasion, loss recovery, and education. They also encounter numerous challenges, such as seniors' refusal of help and considerable mental and financial stress. Drawing on these, we develop a conceptual framework to characterize family support in senior-targeted fraud, and outline implications for researchers and practitioners to consider the broader stakeholder ecosystem and cultural aspects.

\end{abstract}




\maketitle

\section{Introduction}
%-------------------------------------------------------------------------------
Online fraud is a highly prominent topic in the domains of Human-Computer Interaction (HCI) and Security and Privacy (S\&P)~\cite{razaq2021we,mcdonald2023don,wash2020experts,vitak2018knew,6234414,4223213}. \modi{It is increasingly recognized as a critical global issue that impacts communities worldwide~\cite{button2014online,url20,button2012cross,cross2022meeting,url2,url4}.} The elderly might be disproportionately targeted and suffer more severe consequences when falling for online fraud, especially when they experience cognitive decline \cite{james2014correlates,han2016mild}, overly trust others \cite{li2013age,castle2012neural}, are socially isolated \cite{alves2008effects,lachs2015age}, or lack knowledge about fraud prevention \cite{boyle2013cognitive,dadalt2016older}. \modi{In China, 78\% of seniors have received scam messages, with fake online shopping and investment being the common schemes, and between 20-30\% seniors fall for the scam eventually \cite{url22}.}
%% with fraudsters using tactics like fake online shopping and investment schemes that affected 32.4\% and 21.7\% of this demographic respectively 
%,9833665,mendel2019my
%Berry , such as  and regularly checking their bank statements \cite{}. ,
%8835345
%burnes2017prevalence

%According to the US Federal Bureau of Investigation (FBI), Americans lost \$12 billion to online fraud just in 2023~\cite{url2}.

%Senior-targeted online fraud is a global issue. In 2023, the FBI's Internet Crime Complaint Center (IC3) received over 101k fraud complaints from those aged 60+, resulting in \$3.4 billion in losses \cite{url4}. 

To address this global threat, governments, organizations, and researchers worldwide have invested great efforts in safeguarding seniors from fraud. \modi{For instance, China enacted the Anti-Telecom and Online Fraud Law of the People's Republic of China to prevent, curb, and penalize online fraud activities in 2022 \cite{url48}. China's Ministry of Public Security launched the National Anti-Fraud Center (NAFC) mobile app to detect and report potential telecommunication fraud \cite{url47}.} The US Federal Trade Commission (FTC) has dedicated advisory groups \cite{url20} and Age UK advocated for financial institutions to implement various nudges and frictions in their payment system to safeguard seniors from fraud \cite{url21}.

Prior academic research has highlighted the significant role \textit{family members} play in safeguarding seniors from fraud \cite{deliema2023correlates,das2018breaking,deliema2018elder}, such as how family members are often the primary sources of cybersecurity information for the elderly~\cite{nicholson2019if}. However, existing studies related to senior-targeted fraud have primarily focused on seniors' perspectives, such as risks and threats faced by older adults \cite{kemp2023consumer,carlson2006phishing,soffer2023old} and their diverse needs \cite{parti2023if,button2009better}. Little research has empirically and systematically investigated \textbf{how family members safeguard seniors from online fraud} in practice. 

Focusing on only seniors' perspectives has a potential limitation: underreporting. Prior work has revealed many seniors were hesitant to disclose their experiences of fraud due to self-blame, concerns about family reputation, and other factors \cite{van2001investigating,parti2023if}. Additionally, some reports indicated that seniors might not fully grasp how their behaviors could make them vulnerable to fraud \cite{oliveira2017dissecting}. As such, family members can offer a fresh and supplementary perspective in understanding and combating senior-targeted fraud. 

Moreover, little is known about what family supportive behaviors are taken in order to counter senior-targeted online fraud and what challenges family members face when assisting older adults. Investigating senior-targeted fraud through the lens of family members has the potential to shed light on the diverse scope and comprehensiveness of senior-targeted fraud, provide novel insights into the actual role of family support in anti-fraud efforts~\cite{muralidhar2019rethinking}, and highlight the design, policy, and educational opportunities against senior-targeted fraud as a sociotechnical problem~\cite{shao2019older}.
%we may lack a comprehensive understanding of senior-targeted online fraud
%as a supplementary perspective

% Another study advocated for family members to serve as protectors against senior-targeted fraud such as being alert to warning signs and closely watching elderly relatives' financial activities \cite{berry2013effect}. 

%Answering this question can provide novel insights into the actual role of family support in anti-fraud efforts~\cite{muralidhar2019rethinking} as well as highlight the design, policy, and educational opportunities against senior-targeted fraud as a sociotechnical problem~\cite{shao2019older}. 
% and intertwines with older adults' financial autonomy

In China, the assistance provided by younger family members to the elderly is particularly important. Unlike Western cultures, Chinese society generally views younger family members' support for the elderly as an obligation and responsibility (i.e., filial piety) \cite{xu2001family}. Family is usually the primary source of both financial and caregiving support for older adults in China, making family support crucial in the lives of Chinese seniors \cite{yue1999filial,zhan2011role}. By investigating the support ecosystem against senior-targeted online fraud in China from the lens of family members, we also illuminate the social-cultural aspects of fraud prevention and mitigation. Our research centers around the following questions:

\begin{itemize}
\item \textbf{RQ1}: What vulnerabilities and corresponding senior-targeted online fraud are disclosed by younger family members?  \textbf{(seniors' vulnerabilities and fraud practice)}

\item \textbf{RQ2}: What measures do younger family members take to counter senior-targeted online fraud? \textbf{(family support behaviors)}

\item \textbf{RQ3}: What are the challenges when younger family members attempt to assist seniors? \textbf{(challenges in family support)}

\end{itemize}

To answer these questions, we conducted an inductive thematic analysis of 124 posts within 16,872 comments on RedNote (Xiaohongshu),\footnote{\url{https://www.xiaohongshu.com/explore}} a popular social media platform in China, to explore family support against senior-targeted online fraud. Our findings highlighted how online fraud was crafted to target seniors' vulnerabilities in China, such as by exploiting \textit{seniors' concerns about health and well-being} 
% to \textit{induce the purchase of fraudulent health online courses}), 
and \textit{social relationships} 
% was a common fraud channel 
(RQ1). Younger family members' supportive behaviors occur throughout the entire anti-fraud process, including \textit{prevention}, \textit{identification}, \textit{persuasion}, \textit{loss recovery} and \textit{education} (RQ2). Nonetheless, younger family members also faced challenges in family support, such as \textit{seniors guarding their information} and \textit{refusal of help}, as well as the \textit{enormous mental and financial stress} they experienced in helping seniors recover from the fraud (RQ3). We discussed our findings' implications and recommendations to counter senior-targeted fraud under the family support ecosystem.

Our work makes the following contributions to HCI.

\begin{itemize}

\item We unpack the family support ecosystem against senior-targeted online fraud in China, using the perspectives of younger family members as a novel lens, and show that the support is a long-term and evolving process including various stakeholders.

\item We reflect on how sociocultural factors in China shape the family support ecosystem, leading to tensions between financial autonomy and security for older adults.

\item We demonstrate challenges for younger family members to support the elderly. These challenges inform recommendations of interventions to be deployed by platforms, education for older adults, and the support needed for younger family members themselves.

\end{itemize}

%\item We collect a comprehensive dataset of senior-targeted online fraud reported by younger family members on social media. The dataset serves as a valuable resource for future HCI research in online fraud.

%\cite{10190490,10190508,10190672}
%6234430,8418597,8802443

\section{Related Work}

\subsection{Online Fraud and Anti-fraud}

Online fraud is defined as ``the experience of an individual who has responded through the use of the internet to a dishonest invitation, request, notification or offer by providing personal information or money that has led to a financial or non-financial loss or impact of some kind'' by Cassandra Cross \cite{cross2014challenges}. While there are multiple definitions of online fraud, a fundamental aspect is the use of the internet as a means to perpetrate fraudulent activities \cite{kavrestad2014defining}. The World Economic Forum identifies cyber threats as one of the most significant global risks \cite{url3}.

Fraud and anti-fraud efforts are deeply intertwined. A large body of existing works has examined fraud from the attackers' viewpoint by studying their psychological tactics \cite{schmitt2023digital}, attack intentions \cite{bera2023towards}, and fraudulent techniques \cite{conti2010malicious}. For instance, Mouton et al. introduced a comprehensive framework for social engineering attacks
% , which involved manipulating individuals into revealing confidential information through deceit 
\cite{mouton2016social}; building on this framework, Razaq et al. investigated the different schemes used by fraudsters
% based on the social engineering attack framework 
\cite{razaq2021we}. These endeavors provided valuable insights into anti-fraud measures like fraud detection and prevention systems \cite{abdallah2016fraud, krishna2018evaluating,barse2003synthesizing}. Additionally, other research has focused on the victims' perspective by investigating why individuals fall for scams \cite{button2014online} as well as their perceptions and experiences
% , behaviors, or perceptions, including security and privacy concerns, security information sources, etc. 
\cite{herbert2022digital}. These findings demonstrated victims' anti-fraud strategies and inspired designs for educational and intervention programs \cite{zou2024cross,jansen2020social}. 
%On the application front, numerous fraud detection systems \cite{phua2010comprehensive} and intervention projects \cite{prenzler2020works} have been developed to combat fraud. A notable example is the National Anti-Fraud Center (NAFC), a mobile app launched by China's Ministry of Public Security in March 2021, designed to detect and report potential telecommunication frauds \footnote{\url{https://en.wikipedia.org/wiki/National_Anti-Fraud_Center}}.

However, according to family systems theory \cite{bowen1993family}, individuals often do not operate in isolation, as they share many material and emotional connections with family members. Moreover, many previous works highlighted the \modi{significance} of family members in safeguarding seniors from fraud \cite{deliema2023correlates,deliema2018elder}. As such, our research draws from the perspectives of younger family members to enrich the understanding of senior-targeted online fraud, rather than solely focusing on seniors themselves or fraudsters.
% should not be viewed in isolation but as part of a family unit, which is inherently emotional and interconnected. 

\subsection{Online Fraud and Seniors}
A related study highlighted that an average of 9.1\% of adults aged 55 to 64, 7.3\% of those between 65 and 74, and 6.5\% of individuals over 75 fell prey to consumer fraud in the US \cite{url5}. There are numerous challenges in protecting the elderly against online fraud. 

First, factors such as cognitive decline, an overly trusting nature, social isolation, and a lack of knowledge regarding fraud prevention can heighten seniors' vulnerability to online fraud
%psychological vulnerability,
% seniors show a high level of vulnerability to fraud 
\cite{shao2019older,lichtenberg2016psychological,lichtenberg2013psychological,carstensen2006influence,carstensen2005intersection,kyi2022don}. Second, there is a proliferation of scams specifically targeting older adults. The National Council on Aging in the US reported the top five financial scams aimed at older adults, such as government impersonation scams and grandparent scams \cite{url1}. Third, mainstream security mechanisms and application designs often do not adequately consider the needs and perspectives of older adults. For example, older adults might struggle with managing passwords \cite{hornung2017navigating}, which may amplify potential senior-targeted fraud. Lastly, some older adults may exhibit lower proactive tendencies in avoiding fraud risks. Frik et al. indicated that older adults commonly employed passive strategies, such as limiting or avoiding technology use, while active mitigation strategies were less common \cite{frik2019privacy}. However, Zou et al. found that the elderly applied some active strategies, such as configuring privacy and authentication settings, and selectively disclosing sensitive information \cite{zou2024cross}.

%psychological vulnerability \cite{lichtenberg2016psychological,lichtenberg2013psychological} changes in emotional regulation \cite{carstensen2006influence,carstensen2005intersection,charles2008unpleasant}, risk-taking behavior \cite{holtfreter2005consumer,denburg2007orbitofrontal},
%Nevertheless, it is not just age that makes someone more likely to be scammed. Several studies suggested that other aspects, like gender, income, risk perceptions could also play a role \cite{vitak2018knew,oliveira2017dissecting}. and healthcare technologies could pose privacy and security concerns \cite{frik2019privacy}
Furthermore, the harm caused by senior-targeted fraud is profound. The most apparent impact is financial loss.
% , as previously mentioned. 
Moreover, falling victim to fraud often leads to traumatic experiences with significant physical and mental health repercussions. Victims might suffer from major depression, generalized anxiety disorder, feelings of anger, shame, regret, and various somatic symptoms \cite{button2014not,ganzini1990prevalence}. The repercussions extend beyond seniors to their families, who may be equally affected by the financial and emotional consequences \cite{berry2013effect}. In essence, senior-targeted fraud is a pervasive, challenging to combat, and highly detrimental issue. 

What makes anti-fraud efforts harder is the under-reporting among older scam victims. While the majority of scams go unreported \cite{button2014not}, the reporting rate is even lower when it concerns older adults \cite{van2001investigating, url6}. Previous work has shown that senior victims were not willing to report crimes or seek assistance from their family or community due to a lack of accessible or sufficient emotional, educational, and technical support \cite{parti2023if}. Consequently, it is highly likely that the actual incidence and prevalence of senior-targeted fraud extend far beyond the reported ones. 

Prior research on senior-targeted fraud has largely focused on the seniors' perspectives through interviews \cite{segal2021consumer,reisig2013shopping} and surveys \cite{alves2008effects,judges2017role}, etc. However, limited research is from the viewpoint of those close to seniors and aware of their online activities, i.e., \textit{younger family members}, who may provide complementary
% relatively objective and comprehensive 
insights. For example, for the under-reporting issue \cite{parti2023if}, the disclosures from family members may shed light on the diverse scope of senior-targeted fraud. Moreover, older and younger individuals exhibit different patterns in detecting lies and discerning trustworthy faces in offline contexts~\cite{ruffman2012age,castle2012neural}, which could carry over to the online fraud context. Our work aims to address this gap by exploring the nuances of seniors' vulnerabilities, the various types and channels of fraud, and the tactics used by scammers through the lens of younger family members (RQ1).

%,oliveira2017dissecting

\subsection{\modi{Family Dynamics in Safeguarding Seniors from Fraud}}

Research on fraud against the elderly has increasingly underscored the significance of extending anti-fraud efforts from seniors to their family members \cite{deliema2023correlates,parti2023if}. For example, DeLiema encouraged the involvement of family members in financial decision-making processes and the implementation of legal and financial safeguards to protect the assets of older adults \cite{deliema2018elder}. Scammers also noticed the potentially protective role of family members in countering fraud, so they exploited social isolation to manipulate targets to hide information from family members \cite{deliema2018elder,deliema2023correlates}. In addition to the benefits that come with family involvement, there might be tensions and challenges during the support process. Younger family members might become overly paternalistic in their efforts to protect older adults, leading to what is known as ``family surveillance'' \cite{murthy2021individually}. \modi{Moreover, Parti et al. demonstrated that in the realm of online fraud, US older adults were often hesitant to disclose their experiences of fraud to their family members due to self-blame, concerns about losing family status, particularly among younger family members, and trust issues \cite{parti2023if}}. 

%This reluctance to share information with family members potentially exacerbates the challenges faced by family members in providing support.

%unlike in general cybersecurity contexts where older adults are willing to rely on younger family members for support \cite{schreuers2017problematizing,herbert2022digital, murthy2021individually}, 

%Even with good intentions, younger family members' monitoring and protective actions could limit the autonomy of older adults and impede their ability to learn about digital threats and self-protection skills \cite{murthy2021individually}.
%In fact, fraud committed by a family member is a common form of elder financial abuse \cite{url7}.
%herbert2022digital
%,zou2024cross
%Moreover, older adults generally acknowledged their limitations in terms of digital literacy and relied on their family members for cybersecurity support.

%In the realm of digital security, 
%Initially, family members might provide helpful advice on financial matters, but family members would gradually assume complete control over seniors' affairs and misappropriate funds for personal gain \cite{smith2000fraud}. 

%Identifying the exact point at which abuse occurred could be challenging, due to unclear circumstances and difficult evidence collection \cite{smith2000fraud}.

The cultural, economic, and technological contexts in China have a profound influence on the family's role in anti-fraud efforts. The high regard for filial piety within Chinese society shapes the cultural norm that the younger generation should care for and protect their elders \cite{xu2001family}. Additionally, traditional parenting and family relationships in Chinese families give parents an authoritative or authoritarian position \cite{chuang2018parenting}. This internal power structure within families might pose obstacles for the younger generation to intervene and prevent fraud. Economically, younger family members often offer financial care to parents within Chinese families \cite{yue1999filial,sheng2006intergenerational}. Unlike in the US, where government assistance such as Medicaid is available for elderly adults who have exhausted their funds for long-term care, in China, family members represent the last line of financial security \cite{zhan2011role}. In this context, the financial support and education provided by the younger family members are crucial for preventing senior-targeted fraud. 

% The risk of online fraud is heightened for the less tech-savvy elderly population, underscoring the importance of younger generations in guiding and protecting their elders in the digital realm.

%Furthermore, China's digital payment landscape, dominated by services like WeChat Pay and Alipay, significantly differed from other countries, with over 90\% of mobile payment users utilizing these platforms \cite{he2023have,shen2020can}. 
%

%Despite the crucial role played by younger family members in safeguarding seniors from fraud, how they exactly protect seniors against online fraud has rarely been investigated. Our research examines the family support ecosystem by shedding light on younger family members' supportive behaviors (RQ2) and challenges (RQ3).

%To the best of our knowledge, this study is the first to comprehensively investigate the roles and challenges faced by younger family members in supporting seniors against online fraud.

\modi{Prior research has examined family support in the context of older adults' technology use and finances, such as family's support in financial technology onboarding~\cite{tang2022never}, family's account sharing for digital inclusion~\cite{he2023have} and general online security and privacy support \cite{murthy2021individually}. Our study extends this work by shedding light on family support in safeguarding seniors from online fraud. While existing research on senior-targeted fraud often mentions the significance of younger family members, they primarily focus on the perspectives of the seniors themselves \cite{deliema2018elder,kemp2023consumer,parti2023if} without deeply investigating how younger family members engage in online fraud safeguarding. Moreover, family support in senior-targeted fraud within the non-Western context of China remains unexplored. Our research aims to fill the research gap by examining the family support ecosystem by exploring younger family members’ supportive behaviors (RQ2) and challenges (RQ3).}
%Much previous research has looked into how family members support older adults' technology use and finances, such as family's support in financial technology onboarding~\cite{tang2022never} and family's account sharing for digital inclusion~\cite{he2023have}. Our study further extends this line of work to family's safeguarding of older adults' online financial security. Moreover, 
%This reliance on family members for support also carries certain risks, especially when it involves sharing sensitive digital resources \cite{watson2020we}. 
\section{Method}
\begin{figure}[htbp]
	\centering
	{\includegraphics[width=1\columnwidth]{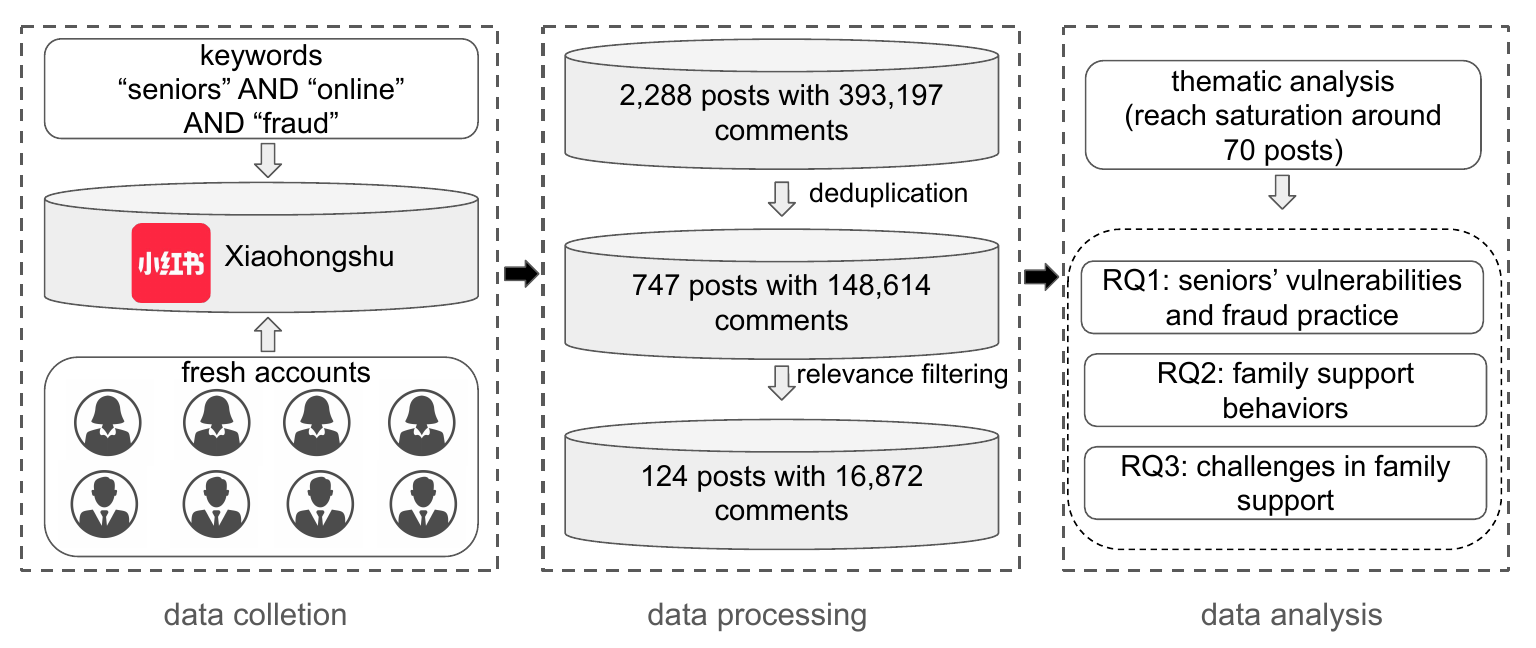}}\\
  \caption{The analytical flow to understand younger family members' safeguarding against senior-targeted fraud.}
  \Description{Figure 1: We crawl data on RedNote based on the keywords ("seniors" AND "online" AND "fraud") and eight fresh accounts. We got 2,288 posts with 393,197 comments, and after deduplication and relevance filtering, we obtained 124 posts and 16,872 comments. Then we utilized thematic analysis to find seniors' vulnerabilities, family support behaviors and challenges in family support.}
  \label{FIG: analytical flow}
\end{figure}

To systematically understand how younger family members safeguarded seniors from fraud, we collected discussions related to seniors' online fraud on social media and conducted thematic analysis \cite{braun2006using} to discover meaningful patterns about online fraud targeted seniors perceived by younger family members (RQ1), their support behaviors (RQ2), and challenges (RQ3). The overall analytical flow is shown in Figure \ref{FIG: analytical flow}.

\subsection{Data Source}

% To comprehensively understand how younger family members safeguard against senior-targeted fraud, 

We chose social media platforms as our data source, where younger family members shared their experiences about senior-targeted fraud. This method provides the following advantages: (1) Discussions there occurred in a natural setting, unlike in interviews or surveys where responses might be influenced by social desirability bias and the Hawthorne effect \cite{mccambridge2014systematic}. (2) Social media platforms provide access to extensive data, which addresses the limitations of small sample sizes that are typical in qualitative studies. (3) The interactive features on social media (e.g., comments, likes, and shares) can foster ongoing discussions that yield deeper insights.

%We chose to gather data from the perspective of young people. We adopted this method because: (1) For online fraud targeting seniors (RQ1), young individuals from a third-party view . 
%(2) For younger family members' supportive behaviors (RQ2) and challenges (RQ3), no one understood the support provided and the challenges faced by young people better than themselves. Therefore, obtaining insights from young individuals might provide a more comprehensive understanding of the online fraud landscape affecting seniors, and the supportive behaviors and challenges encountered by younger family members.

%To determine the suitable data source, we first conducted preliminary interviews with seniors about their experiences with online fraud. These interviews revealed that seniors might withhold information due to various reasons such as embarrassment or shame, which also aligned with previous studies \cite{parti2023if}. Young individuals from a third-party view, had the potential to provide a more comprehensive understanding of the online fraud landscape affecting seniors.

\begin{figure}[htbp]
	\centering
	{\includegraphics[width=1\columnwidth]{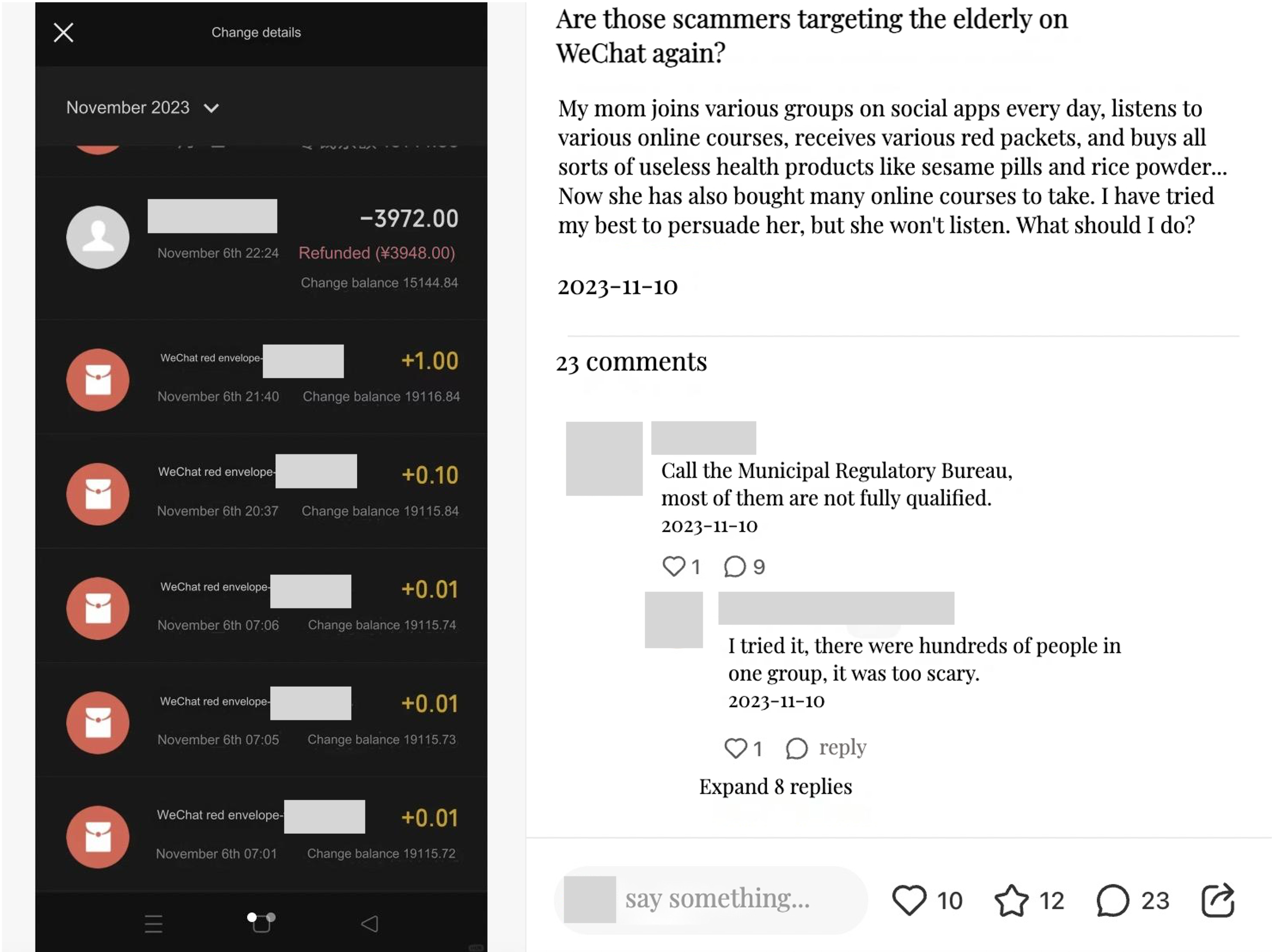}}\\
  \caption{An interface example of RedNote (translated from Chinese).}
    \Description{Figure 2: We gave an online fraud screenshot in RedNote including the transfer records (left) and fraud case describing a younger family member saying how his mother was deceived online and the corresponding comments (right).}
  \label{FIG: interface}
\end{figure}

We focused on discussions related to senior-targeted online fraud on RedNote (Xiaohongshu). RedNote, often referred to as the Chinese Instagram, is a Chinese life-sharing platform \cite{url10}. Users of RedNote often share personal life experiences and opinions through text, photos, and videos, which aligns with our everyday context of senior-targeted online fraud. By 2020, RedNote had over 450 million registered users, and more than 121 million monthly active users \cite{url11,url12}. Many prior work has explored user-generated content on experience sharing on RedNote \cite{fan2023study, cai2022changing,tao2023social}. We also looked through the related content on other popular social media platforms in China such as Weibo \cite{url23} and WeChat \cite{url24}, but content about senior-targeted online fraud on these platforms tends to be more focused on news and official advice rather than personal stories. RedNote, similar to Instagram, has a threaded discussion structure where each initial post is followed by a series of subsequent comments over time. This structure allows for the observation of discourse and interaction dynamics among users. A screenshot of RedNote's interface 
% about senior-targeted online fraud disclosed by younger family members 
is shown in Figure \ref{FIG: interface}.

%Furthermore, the user demographics of RedNote were predominantly young, with more than 90\% of its users under the age of 44, fitting our objective of investigating this context from the perspective of younger individuals. 
\subsection{Data Collection}

Before collecting data, we studied the search mechanism of RedNote. We discovered that searching a keyword does not only bring up content containing that exact keyword but also surfaces content related to the keyword, suggesting that RedNote's search algorithm returns posts based on contextual similarity rather than specific queried words. For our study, the population featured was seniors, our context was online environments, and our focus was fraud. Therefore, we set our search keywords to ``seniors'' AND ``online'' AND ``fraud'' to gather content pertinent to our research topic. We employed data crawling assisted by MediaCrawler tool based on these keywords \cite{url8}. The search also yielded posts associated with ``parents,'' ``grandparents,'' ``mobile,'' ``internet'' and other words related to our keywords, which ensured the richness of our data.

Furthermore, RedNote has a personalized content recommendation algorithm that tracks users' interactions, such as clicks, viewing time, and likes, to tailor content recommendations \cite{wu2018auditing, setyani2019exploring}. Therefore, to mitigate the influence of previous usage records on our data collection, we created unused accounts for crawling data, a common practice recommended in previous research on crawling social media data \cite{gjoka2011practical,ng2022not}. When creating new accounts on RedNote, the selection of gender and age could also impact the recommended content. To ensure a diverse data collection, we established eight different accounts, corresponding to male and female users aged 21, 30, 40, and 52 years. These ages were based on median values derived from the age distribution of RedNote's adult user profile \cite{url13}. \modi{We did not select older age for user accounts, considering the possibility of older members being victims themselves, which may introduce noise to the dataset.} We adopted the default search sort of RedNote and collected all types of data including texts, videos, and images. Each search returned around 200 posts. The limited size of search results on RedNote prevented us from creating a comprehensive dataset. Nonetheless, the search mechanism of RedNote based on contextual similarity and our collection strategies with diversified identities ensured the representativeness and randomness of
collected data. Ultimately, we completed our data collection in January 2024, and we retrieved 2,288 posts with 393,197 comments.
%We did not select older age for user accounts considering the typical user demographic of the platform, where most users are young people and only 8.2\% of users are over the age of 44 \cite{url13}. Additionally, older members (e.g., more than 60) as younger family members are themselves likely victims of senior-targeted fraud.

\subsection{Data Processing}
After removing duplicated cases, we had a collection of 747 posts with 148,614 comments. However, the dataset, even after deduplication, still contained some unrelated entries, such as anti-fraud advertisements, public awareness campaigns, and personal stories of being scammed. 

To filter out these irrelevant contents, the first and second authors independently coded all the posts with the inclusion criteria that focused on data related to younger family members disclosing experiences of seniors being deceived. The two coders reached high inter-rater reliability (Cohen's kappa$=$$0.91$). Following this, the two coders engaged in discussions to resolve any discrepancies and reach a consensus on all the annotations. After the filtering process, our final dataset contained 124 posts with 16,872 comments ranging from May 2020 to January 2024 \cite{kou2024trading,li2023s}, specifically related to the disclosures by younger family members about the online fraud experiences of seniors. We also collected user profiles for each user, including user ID, gender, IP locations, and age. These posts were created by 124 unique users and comments were created by 12,383 unique users. The users' IP locations covered all provinces in China. The median age was 26 and the average age was 28, with the 25th percentile at 22 and the 75th percentile at 32.\footnote{We found only 3,161 users (25.47\%) disclosed their ages, and some reported unrealistic ages such as 2023 years old. After removing such outliers, we got the statistical data of ages. But note that our data analysis was based on the content of the posts and comments rather than on age, and the ages listed in profiles might be fabricated \cite{magdy2017fake}. Therefore, these ages should not be considered as accurate representations of users' actual ages but rather as reference data, which we have acknowledged in limitations.}
%To ensure the reliability of our coding process, we used Cohen's Kappa \cite{mchugh2012interrater}, which is a statistical measure for assessing inter-rater reliability in qualitative research.This measure was generally considered more robust than simple percent agreement because it accounted for the possibility of chance agreement.
\subsection{Data Analysis}

We applied inductive thematic analysis \cite{braun2006using}, a qualitative method that entails searching across a dataset to identify, examine, and report repeated patterns, to thoroughly explore three research questions focusing on seniors' vulnerabilities and online fraud practices (RQ1), the supportive behaviors of younger family members (RQ2), and their challenges (RQ3). Initially, two authors independently performed open coding \cite{corbin2014basics} on a random sample of 50 posts and their comments to generate preliminary codes (e.g., \textit{the elderly did not tell the scammed amount''} and \textit{``arguments with seniors''} related to RQ3 about challenges). This stage directly reflected insights derived from the data. Subsequently, we discussed and reconciled our codes to ensure a unified understanding and constructed a codebook. Then we evaluated our codebook by applying it to additional posts and comments and documenting newly merged codes. We repeated the process, where we revisited and refined our codebook in weekly meetings. This iterative coding process continued until saturation was reached when approximately 70 posts with more than 12,000 comments were coded. Then axial coding \cite{corbin2014basics} was employed to identify connections between these codes, fostering the development of broader themes. For example, the codes \textit{``the elderly did not tell the scammed amount''} and \textit{``younger family members did not know the fraud''} were linked under a broader theme of \textit{``guarding information.''} This thematic connection indicated a behavior pattern among seniors regarding their experiences with fraud. 
%Since our primary goal was to generate emerging codes and themes during data analysis, we did not calculate the inter-rater reliability (IRR) to seek theoretical agreement \cite{mcdonald2019reliability}.

%By merging similar codes, we systematically categorized them into xx themes and xx sub-themes.

% \subsection{Researcher Positionality}
% The two coders were born and raised in China and all authors have research experience about older adults' financial security.
\subsection{Ethics Statement} Our study received approval from our Institutional Review Board (IRB). When collecting and analyzing users' public data for research purposes, it's still essential to carefully consider the potential risks, benefits, and harm involved \cite{fiesler2019ethical}. We believe this research provides significant benefits by highlighting seniors' vulnerabilities to fraud and fraud practice, exploring family support behaviors, and identifying challenges in providing support. Our intention is not to expose individual fraud experiences or single supportive behavior and challenge but to uncover broader patterns in collective online behaviors. To minimize the risk of linking quotations back to specific users, we implemented several measures. First, we removed all usernames when presenting data. Second, all quotes were translated from Chinese to English using Google Translate and then paraphrased to further reduce searchability. The final translations were reviewed and refined by the first author to ensure accuracy. Moreover, collecting data from the perspective of younger family members on social media avoids direct contact with seniors and their families, thereby preventing the ethical concerns of retraumatizing them by recounting distressing experiences \cite{connolly2010elder,consalter2024technology}.

\section{Findings}
Our findings contribute a rich understanding of how younger family members safeguard seniors from online fraud. We first report on seniors' vulnerabilities and corresponding senior-targeted online fraud disclosed by younger family members in \S\ref{finding: RQ1} (RQ1). We then demonstrate the supportive behaviors of younger family members in \S \ref{finding: RQ2} (RQ2). We finally reveal the challenges faced by younger family members during their assistance process in \S\ref{findings: RQ3} (RQ3).

\subsection{RQ1: Senior-Targeted Online Fraud}
\label{finding: RQ1}
Our work added more nuances to
prior research that systematically examined online senior-targeted fraud, including the vulnerabilities targeted, fraud types, fraud channels, and fraudsters’ strategies. Our findings were drawn from the perspectives of younger individuals, but this did not imply that their viewpoints were definitive. Previous work also indicated perceptual differences between seniors and young regarding lies and trust \cite{ruffman2012age,castle2012neural}. Instead, the perspectives of younger individuals provide a new lens for understanding the complex nature of elderly fraud. The complete taxonomy is shown in Table \ref{table1}.
\begin{table*}
	\caption{Vulnerabilities and online senior-targeted fraud disclosed by younger family members. ``Fraud Channel'' refers to the crucial steps for money outflow, while ``Fraud Strategy'' denotes the tactics employed by fraudsters. * means there are other types of fraud utilizing the same fraud channel or strategy.}
 \Description{Table 1: There are four parts including seniors' vulnerability, fraud type, fraud channel and fraud strategy. Vulnerability includes health and well-being concern, financial expectation, overreliance on authority, social isolation, generational affection and low digital literacy. There are 11 fraud types (such as unnoticed charges), 10 different fraud channels (such as transactions built through social relationships on social apps) and 20 fraud strategies (such as offering rewards for referrals).}
 \small
 \label{table1}
\begin{tabular}{|p{3cm}|p{3cm}|p{4cm}|p{4cm}|}
\hline
\textbf{Vulnerability}                          & \textbf{Fraud Type}                                                            & \textbf{Fraud Channel}                                                               & \textbf{Fraud Strategy}                                                   \\ \hline
\multirow{11}{3cm}{health and well-being concern} & \multirow{5}{3cm}{false advertising of health products}                          & \multirow{3}{4cm}{C1*: transactions built through social relationships on social apps\footnote{\textit{This harm could occur through various means, including both direct money transferring or gifting on social apps \cite{xu2021lucky}, and indirect digital (e.g., guiding to invest on other apps) or physical transactions (e.g., cash on delivery) mediated through private messaging or groups on social apps.}}} & S1*: offering rewards for referrals                                       \\ \cline{4-4} 
                                                &                                                                                &                                                                                      & S2*: leveraging expert endorsements for promotion                         \\ \cline{4-4} 
                                                &                                                                                &                                                                                      & S3*: sending red packets on social app                                    \\ \cline{3-4} 
                                                &                                                                                & \multirow{2}{4.4cm}{C2*: live-streaming purchase links}                                  & S4*: creating benefit illusion (free gifts/insider price)                 \\ \cline{4-4} 
                                                &                                                                                &                                                                                      & S5: amplifying seniors' guilt                                             \\ \cline{2-4} 
                                                & \multirow{5}{3cm}{purchase induction of fraudulent health and lifestyle courses} & \multirow{3}{4cm}{C1*: transactions built through social relationships on social apps} & S2*: leveraging expert endorsements for promotion                         \\ \cline{4-4} 
                                                &                                                                                &                                                                                      & S3*: sending red packets on social app                                    \\ \cline{4-4} 
                                                &                                                                                &                                                                                      & S4*: creating benefit illusion (free trial/insider price)                 \\ \cline{3-4} 
                                                &                                                                                & \multirow{2}{4.4cm}{C2*: live-streaming purchase links}                                  & S6: affirming seniors                                                     \\ \cline{4-4} 
                                                &                                                                                &                                                                                      & S7: disparaging medicine for self-promotion                               \\ \cline{2-4} 
                                                & purchase induction based on superstition                                    & C3: live-streaming virtual gifts                                                     & S8: blessing for families                                                 \\ \hline
\multirow{9}{*}{financial expectation}          & \multirow{5}{*}{investment}                                                    & \multirow{2}{*}{C4: investment app/website}                                          & S1*: offering rewards for referrals                                       \\ \cline{4-4} 
                                                &                                                                                &                                                                                      & S2*: leveraging expert endorsements for promotion                         \\ \cline{3-4} 
                                                &                                                                                & \multirow{3}{4cm}{C1*: transactions built through social relationships on social apps} & S9*: luring with high profits                                             \\ \cline{4-4} 
                                                &                                                                                &                                                                                      & S10: pretending to be helpless                                            \\ \cline{4-4} 
                                                &                                                                                &                                                                                      & S11: disguise the project as national                                     \\ \cline{2-4} 
                                                & \multirow{2}{3cm}{false advertising of collectibles}                             & \multirow{2}{*}{C2*: live-streaming purchase links}                                  & S12: highlighting collectible value                                       \\ \cline{4-4} 
                                                &                                                                                &                                                                                      & S13: presenting compelling stories                                        \\ \cline{2-4} 
                                                & \multirow{2}{3cm}{purchase induction of fraudulent money-making courses}         & C1*: transactions built through social relationships on social apps                  & S1*: offering rewards for referrals                                       \\ \cline{3-4} 
                                                &                                                                                & C2*: live-streaming purchase links                                                   & S9*: luring with high profits                                             \\ \hline                   
\multirow{2}{*}{overreliance on authority}      & \multirow{2}{*}{authority impersonation}                                       & \multirow{2}{4.4cm}{C5: transactions instructed through video calls}                     & S14: pretending to be authorities to offer assistance                     \\ \cline{4-4} 
                                                &                                                                                &                                                                                      & S15: educating fraud knowledge to decrease suspicion                      \\ \hline
social isolation                                & romance                                                                        & C1*: transactions built through social relationships on social apps                  & S16: exploiting intimacy to deceive for money or information              \\ \hline
generational affection                          & family member impersonation                                                    & C1*: transactions built through social relationships on social apps                  & S17: pretending to be family members in urgent need of money        
\\ \hline

\multirow{5}{*}{low digital literacy}           & \multirow{4}{*}{unnoticed charges}                                    & C6: password-free payment                                                            & \multirow{2}{4.3cm}{S18: exploiting unintentional enrollment to deduct money} \\ \cline{3-3}
                                                &                                                                                & C7: auto-renewal service subscription                                                &                                                                           \\ \cline{3-4} 
                                                &                                                                                & C8: pop-up ads with fake/transparent close buttons                                   & \multirow{2}{4cm}{S19: popping up red packets to induce clicks}             \\ \cline{3-3}
                                                &                                                                                & C9: fake red packets redirecting to payment pages                                    &                                                                           \\ \cline{2-4} 
                                                & fake payment                                                                   & C10: fraudulent QR code scanning                                                     & S20: falsely claiming to have paid                                        \\ \hline

\end{tabular}
\end{table*}

\textbf{\textit{Vulnerabilities of seniors.}} 
We identified various types of fraud and how they target seniors' vulnerabilities. Seniors' \textbf{concerns for health and well-being} made them easy targets for online frauds involving \textit{false advertising of health products}, \textit{purchases of fraudulent health and lifestyle courses}, and \textit{purchases based on superstition}. Additionally, seniors' \textbf{financial expectations} might lead them to fall prey to \textit{investment fraud}, \textit{false advertising of collectibles}, and \textit{purchases of fraudulent money-making courses}. \textbf{Overreliance on authority} could be exploited by scammers \textit{impersonating authorities}. Furthermore, \textbf{social isolation} might open doors for \textit{romance fraud}, and \textbf{caring about the younger generations} could sometimes be manipulated by \textit{family member impersonation fraud}. Moreover, \textbf{low digital literacy} potentially hindered seniors from recognizing signs of fraud. For instance, seniors might \textit{not notice charges deducted from their bank accounts}. Some seniors working as street vendors could also be deceived by \textit{fake payments} when they were told the money had been paid, but it was not. Each of these vulnerabilities contributed to the risk profile of seniors and put them in a dire situation when online fraudsters also use sophisticated channels and strategies.

\textbf{\textit{Unrecognized fraud schemes among seniors.}}
By examining online fraud from the perspectives of younger family members, our research revealed some frauds that seniors might overlook, demonstrating the critical role of intergenerational insights in protecting seniors from online fraud. For instance, seniors were often enticed by the fabricated but compelling stories presented in live-streaming, which led them to believe in the high value of collectibles. As a younger family member pointed out,
\begin{quote}
    \textit{``My elderly family members got hooked on live-streams and spent a lot of money buying a bunch of so-called antiques, paintings, gold jewelry \dots We even took these items to museums for expert appraisal, only to find out that they were all fake. But they still have unwavering trust in the live-stream host.''}
\end{quote}

Another example was \textit{unnoticed charges}. These often remained overlooked until younger family members noticed unusual charges on the seniors' bills. Some online short videos or novels require payment to continue viewing after an initial free portion. Seniors, sometimes due to low digital literacy or password-free payment settings, would inadvertently make money transfers while engaging with these platforms, not fully realizing the financial commitments involved. As one younger family member explained, 
\begin{quote}
\textit{``My dad was maliciously deducted a lot of money from reading novels before, 100 yuan at a time. I asked my dad and he said that all the novels he reads were free, and he did not know about the deductions.''}
\end{quote}

\textbf{\textit{Diversity of fraud channels and exploitation of social relationships.}} Fraudsters used various channels to target seniors. Some channels were straightforward, such as \textit{purchase links in live-streaming} and \textit{fraudulent QR codes} to be scanned in physical electronic payments. Other tactics were more deceptive and harder to notice, such as \textit{fake pop-up ads} designed to trick users into making unwanted purchases. Particularly, \textit{the exploitation of social relationships on social apps} served as a breeding ground for numerous fraud types. In one case about \textit{fraudulent money-making courses}, a younger family member found that a salesperson added her mother on WeChat (a popular social app in China) to trick her mother into purchasing courses.

\begin{quote}
\textit{``I discovered my mom was tricked into buying courses worth thousands of yuan. The salesperson who added her on WeChat kept promising that people in their 60s could earn a thousand yuan a day. He constantly pretended to care and showered my mom with attention.''} 
\end{quote}

Additionally, scammers even utilized the calling feature in social apps to establish stronger connections with the elderly. Another young person shared, \textit{``The teacher called my dad on WeChat every day to ask him to attend class.''}

\textbf{\textit{Common strategies used by fraudsters.}} Although the vulnerabilities and types of fraud varied, there were common strategies employed by scammers. For example, the strategy of \textit{offering rewards for referrals} was utilized in three kinds of fraud types. Furthermore, for \textit{false advertising of health products} and \textit{fraudulent health and lifestyle courses}, scammers often sent small amounts of money in \textit{red packets} on social apps to increase the engagement of elderly individuals. \textit{Red packets} on social apps were based on the Chinese tradition of lucky money, where money could be given to others as a gift \cite{url9}. These strategies worked out well when the seniors targeted were too trusting
% highlighted the potential common susceptibilities among seniors, like 
or had a strong desire for small gains. Furthermore, scammers sometimes crafted their strategies to exploit seniors' vulnerabilities specifically. For seniors with an \textit{overreliance on authority}, scammers \textit{posed as officials offering help}. Exploiting \textit{seniors' care for the younger generations}, fraudsters might impersonate younger family members to request money. 
% \usepackage{multirow}
% There were various dark patterns based on online \textit{red packets}, such as \textit{fake, click-baiting red packets} that redirected the user to payment pages
\subsection{RQ2: Supportive Behaviors}
\label{finding: RQ2}
In the section, we show the family support behaviors from the aspects of \textit{prevention} (\S\ref{prevention behavior}), \textit{identification} (\S\ref{identification}), \textit{persuasion} (\S \ref{persuasion}), \textit{loss recovery} (\S\ref{Loss Recovery}), and \textit{education} (\S\ref{Education: Active Involvement in Fraud Education}).

\subsubsection{Prevention: The Involvement of Controlling Behaviors}
\label{prevention behavior}

To safeguard seniors from online fraud, younger family members used a diverse set of preventive measures. These included \textit{restricting online payments}, \textit{managing tasks for seniors}, \textit{monitoring}, \textit{selective blocking} and \textit{setting up anti-fraud defenses}. The common theme was that younger individuals tried to control seniors' online activities to mitigate the risk of fraud.

\textbf{\textit{Online payment restriction.}} Some younger family members restricted online payment of seniors, such as restricting online spending and not allowing seniors to link their bank cards to mobile phones. As a younger family member explained, \textit{{``We do not allow them to link cards to mobile phones and only give them cash.''}}

\textbf{\textit{Managing tasks for seniors.}}
Younger family members would take on the responsibility of performing online tasks on behalf of the seniors, such as registering online accounts and doing online shopping. A daughter said,
\begin{quote}
\textit{{``My mom always asks me to teach her how to link her bank card for payments, but I refuse and tell her to let me know what she needs and I'll purchase it for her.''}}
\end{quote}

\textbf{\textit{Monitoring.}}
% This behavior indicated that younger family members aimed 
Another strategy employed by younger family members is
to monitor seniors' online activities and be prepared to eliminate any risks as soon as they detect anomalies. As one younger family member recalled,
\begin{quote}
\textit{{``I directly logged into my dad's Kuaishou account (a live-streaming platform). When I saw something strange, I immediately refunded items without waiting for delivery.''}}
\end{quote}

\textbf{\textit{Selective blocking.}} Some younger individuals employed technical methods to limit seniors' access to specific applications. For example, one person provided suggestions on how to block certain applications,
\textit{{``Hey guys, you can log in to the router and block access to TikTok and TikTok live-streaming in our home. You can find the domain names online for reference.''}}

\textbf{\textit{Setting up anti-fraud defenses.}} Some younger family members might assist the elderly in setting up patterns on mobile phones or downloading anti-fraud applications. A younger individual told her setting,
\textit{{``Set the seniors' phones to reject calls from unknown numbers. Disable features that allow strangers to like, follow, or send private messages.''}}

Most of these preventative measures involved controlling behaviors. While these approaches might enhance online security for seniors, they also introduced issues, such as conflicts between seniors' autonomy and the control exercised by younger family members, as we elaborated in \S\ref{information segregation}.

\subsubsection{Identification: A Three-Phase Process}
\label{identification}
An essential component of younger family members' efforts to protect seniors was to suspect and confirm incidents of fraud. This identification typically unfolded in a three-stage process, including \textit{initial contacts}, \textit{collecting evidence}, and \textit{making the judgment}.

\textbf{\textit{Stage I: initial contacts.}} This referred to the stage where younger individuals first became aware of the scam, which might be different from when the elderly person first encountered it. For example, seniors might invite their children to participate in some supposedly profitable activities, and it was during this participation that the younger individuals recognized the fraudulent nature of these activities. This corresponded to fraud strategies like \textit{S4: offering rewards for referrals} in Table \ref{table1}, i.e., when fraudsters claim that seniors could receive red packets for adding more people to a group on social apps. Additionally, younger individuals sometimes caught a clue when seniors asked for help with their financial accounts. One young person mentioned, 
\begin{quote}
    \textit{``My mom asked me to look at her WeChat account because she thought there was something wrong. When I checked, I found out she signed up for an online course that cost 2900 yuan.''}
\end{quote}
There were also instances where potential risks were identified through prevention behaviors like \textit{monitoring}, allowing younger family members to intervene and prevent further development of the scam.

\textbf{\textit{Stage II: collecting evidence.}} In this phase, younger people used multiple approaches to gather concrete evidence. Some would use their wisdom to obtain information. For instance, they initially affirmed their parents and encouraged them to share more details, as some parents might hesitate to discuss their situations with their children due to a fear of criticism or a desire to save face. One child said, \textit{``My dad really cares about his reputation and would never admit he's been scammed. To get to the bottom of it, I started by affirming him positively. Then he sent me his TikTok account and the links to the classes he's been taking.''} 

In other cases, younger people even joined the group on social apps created by fraudsters to gather firsthand information about the ongoing scams. For example, a young person suggested to other young people, \textit{``If the elderly ask us to join the group again, we must be proactive and report it to prevent seniors from getting deeper and deeper.''} Except for taking evidence by strategy, some utilized a direct approach, such as examining their parents' financial footprints by checking their payment histories on mobile devices. 

\textbf{\textit{Stage III: make the judgment}.} After collecting evidence, younger family members might synthesize the evidence to make a judgment. They usually evaluated the legitimacy of the business model, 
% professionalism, 
identified obvious misinformation, and compared costs with standard market prices. However, sometimes they were not certain whether a case was a scam, so they would turn to the internet to validate their suspicions. Social media platforms provided a useful venue where younger individuals could inquire if others thought it was a scam and obtain detailed information such as similar fraud experiences, the specific amounts lost, and the names of fraudulent organizations.

\subsubsection{Persuasion: Younger Family Members Rack Their Brains to Convince Seniors}
\label{persuasion}
To convince seniors about the dangers of ongoing scams, younger family members developed a set of persuasive strategies. These efforts were crucial in addressing the often stubborn disbelief among seniors who might not easily accept that they were being scammed. By \textit{presenting direct evidence} and \textit{involving third parties}, younger individuals worked diligently to protect their elderly family members from further exploitation.

\textbf{\textit{Presenting the direct evidence.}} To persuade the elderly about ongoing scams, younger family members developed a set of approaches. Typically, they \textit{provided evidence or shared similar cases of fraud} to demonstrate to the seniors that what they were experiencing was indeed a scam. For example, one daughter recounted,
\begin{quote}
    \textit{``I exposed the truth to my mom tonight. The financial company she invested in has been revoked, but she still believes it. I'm really at my wit's end.''}
\end{quote}

\textbf{\textit{Involving third parties.}} Despite their efforts, seniors often remained in denial, insisting they were not deceived. In such cases, younger people resorted to \textit{involve other stakeholders for the persuation}, such as other family members, the police, and broader communities on social media. For instance, a young person leveraged the ``@'' feature to mention seniors in social media posts so that other users could help to persuade this older person,
\begin{quote}
    \textit{User1: @ [one older person], please take a look at this.}

    \textit{User2: ``Auntie, please don't fall for those smooth talkers. My grandparents are fully convinced that the scammers are looking out for them, and now they think the rest of us are the enemies. It's giving me daily headaches and insomnia.''}
\end{quote}

When it comes to communication styles used for persuasion, some tried to understand, resonate, and discuss how to avoid fraud with seniors. Others directly scolded the elderly or even threatened to call the police to prevent them from falling into fraud, especially when more gentle methods failed. While some persuasion efforts successfully protected seniors from further scams, these also had many negative results. Seniors might respond by \textit{guarding information} or \textit{refusal of help and intervention}, and persuasion might also contribute to strained family relationships, which we elaborate in \S\ref{findings: RQ3}.

\subsubsection{Loss Recovery: Dual Burden of Financial Assistance and Emotional Care}
\label{Loss Recovery}
We observed that younger family members frequently engaged in two key types of support: \textit{financial assistance} and \text{emotional care}. This dual approach encompassed both tangible financial interventions to recuperate losses and strategies to support the emotional and psychological well-being of their affected loved ones.

\textbf{\textit{Financial assistance.}} \textit{Helping seniors repay their debts} was a common form of support. For instance, one child shared, 
\begin{quote}
    \textit{``I'm planning to transfer 30,000 yuan to my mom to cover half of her losses. It hurts to see her suffer, and it's equally painful to part with my money.''}
\end{quote}

Another approach was to \textit{help seniors reclaim money lost}. This included filing complaints and reporting fraudulent activities to relevant authorities, as well as processing refunds and returns on behalf of the elderly. 

\textbf{\textit{Emotional care.}}
In addition to \textit{financial assistance}, younger family members also took on the role of soothing the emotions of the elderly. They often \textit{provided more companionship}, as one younger family member mentioned,
    \textit{{``I comfort her every day \dots We talk things through, and it helps her let off some steam.''}}

Moreover, some young people suggested using \textit{white lies} to alleviate the emotional burden on the elderly, e.g., \textit{{``How about just tell mom we've gotten the money back? Maybe get a friend to put on a little show for her. After all, we can always make the money back.''}}

\subsubsection{Education: Active Involvement in Fraud Education}
\label{Education: Active Involvement in Fraud Education}
We found that fraud education provided by younger family members to seniors involves  \textit{raising awareness of specific fraud schemes}, \textit{informing general attitudes} needed to protect oneself against fraud, and \textit{providing practical advice}. This education spanned the entire fraud process, including before, during, and after the fraud occurred.

\textbf{\textit{Raise awareness.}} Younger family members hoped to educate their parents or grandparents to raise their awareness of fraud prevention. In addition to \textit{basic finance knowledge}, they also shared knowledge about \textit{specific instances of fraud} with seniors, often by forwarding fraud cases on social apps or mentioning recent fraud examples. The younger generation themselves also deepened their understanding of fraud when paying attention to these cases. One person commented, \textit{``This is the first time I've heard about this type of fraud. I will quickly inform my parents. Thank you for sharing.''}

An interesting approach to raising awareness involves \textit{simulating fraudsters} to alert seniors to potential fraud. For example, a daughter tested her father, 
\begin{quote}
    {\textit{``I directly messaged them on WeChat, saying, `Dad, lend me some money,' and my father didn't ask any questions and transferred the money.''}}
\end{quote}

\textbf{\textit{Inform general attitudes.}}
Younger individuals also informed seniors about the general attitudes to keep in mind against fraud, such as the \textit{importance of always communicating with the younger generation before decision making}. Moreover, they educated seniors to \textit{establish a value system against greed}, emphasizing principles such as \textit{``there is no such thing as a free lunch''} and \textit{``there is no such thing as a risk-free investment.''} 

\textbf{\textit{Provide practical advice.}} The younger generation engaged in practical instructions as well, such as teaching their parents about \textit{how to resist or ignore scammers} in certain situations. As a young individual shared, 
    {\textit{``I make it clear to my parents that no matter who calls and asks them to perform any actions related to their bank cards, they should ignore them completely. Especially when they receive messages requiring them to reply for subscriptions or click on websites for information, they should disregard them entirely.''}}
%teaching older adults \textit{search skills}. For instance, a young person shared,  {\textit{``My mom enjoyed watching short drama, so I taught her how to search for free complete series on a specific platform.''}}Furthermore, they provided 
\subsection{RQ3: Challenges in Providing Support}\label{findings: RQ3}

From \S\ref{finding: RQ2}, we observed younger individuals implemented various supportive measures to safeguard seniors from fraud. However, they also encountered various challenges during the assistance process. First, there was resistance from seniors towards the help provided by younger family members as seniors \textit{guard their information} and \textit{refuse to get help}, as shown in \S\ref{information segregation} and \S\ref{intervention refusal}. Second, \S\ref{tension} describes \textit{dark patterns in online payment systems}, which potentially exacerbated the likelihood of seniors falling victim to scams. Furthermore, \S\ref{difficult rights protection} demonstrates \textit{a lack of consumer protection} when younger people advocate for seniors' losses. Finally, \S\ref{mental stress and family strain} documents the \textit{mental and financial stress as well as strained family relationships} experienced by younger family members.

\subsubsection{Guarding Information}
\label{information segregation}
Although younger family members made lots of efforts to assist seniors from fraud, we found that the elderly did not always accept their help and resisted in their own ways, echoing existing studies \cite{parti2023if}. In this section, we described the practice of \textit{guarding information} among seniors. The phenomenon of seniors holding back critical fraud-related information with younger family members permeated various supportive behaviors. 
%including \textit{refusing anti-fraud information} and \textit{withholding information} from the young.

% In terms of \textit{identification}, 

Seniors might prevent young people from accessing suspicious information from the \textit{initial contact phase}. For example, one young person shared, 

\begin{quote}
\textit{``From the time she signed up for the courses until now, I had no idea. She didn't talk about it, not just with me, but also didn't mention it to my dad or anyone else, until I came home for winter break and checked her phone.''} 
\end{quote}

Furthermore, even when younger individuals noticed something unusual, they often encountered resistance from the elderly during the \textit{evidence collection phase}. One person recounted, 

\begin{quote}
    \textit{``My mom started following some I Ching courses and secretly paid a lot of money, not telling any of us. When I asked her, she was as reluctant \dots to tell me anything. She wouldn't say how much she spent or what exactly she learned. When pressed further, she would just dismiss it with `Oh, it's not what you think.'''}
\end{quote}

When younger family members saught to persuade seniors about the fraud's existence, some seniors might deceitfully reassure their children that they understood the warnings and would not fall for scams again, only to secretly maintain contact with scammers. One child expressed profound regret, 

\begin{quote}
    \textit{``I sent countless messages and made phone calls, warning her she'd been scammed and needed to stop the loss in time. But she never told the truth, while claiming that she understood and I had let my guard down. Now, looking back, I should stayed home and watched over her. I deeply regret it.''}
\end{quote} 
Moreover, our findings in \S\ref{persuasion} about persuasive behaviors highlighted that some younger people opted for an intense communication style, like scolding. This might lead to parents withholding information after being scolded, potentially increasing the risk of being deceived multiple times. For example, one person shared, \textit{``After my mother was scammed for over 10,000 yuan, I scolded her, and now she keeps everything hidden from us.''}

Young people also speculated potential reasons for seniors' guarding of information. The first might be a fear of being blamed. Many elderly were afraid of being reprimanded, leading them to withhold information about scams. A granddaughter revealed, \textit{``My grandmother didn't dare tell anyone at home and only told me when I came home for a vacation after she lost several hundred.''} Some elderly individuals might lie about the amount they were scammed to avoid criticism, e.g., \textit{``My mother-in-law had actually invested over 20,000 yuan but lied to us that she only put in 1,000 yuan.''}

Another reason could be seniors' regard of their own superiority. Some seniors believed they were more knowledgeable or experienced than their children and completely disregarded their advice. One child expressed, \textit{``They always say that they have lived more years than us, so they know better. They are not willing to listen to you because they think you lack experience.''} 

Lastly, manipulations by scammers might also play a role in seniors' guarding of information. Scammers preemptively instructed elderly individuals not to discuss their activities with their children, brainwashing them to trust only themselves. One younger individual shared, \textit{``After telling my mother that she was being deceived, she responded, `my teacher said not to tell the children and to believe in my own efforts!'''} Seniors' guarding of information might also explain why some younger family members resorted to more forceful measures, such as \textit{monitoring} in \S\ref{prevention behavior}, to prevent online fraud among the elderly.

%设置了拒加好友 设置成青少年模式也不行 还有什么办法？ 问老人是怎么加上的她也不说 还生气了

%你们是崇洋媚外我爸就是这样，脑子已经被洗没了，骗子会离间爸妈与子女的关系啊

%This pattern of behavior highlights a significant barrier to communication and intervention, as seniors may hold back critical information .

\subsubsection{Refusal of Help and Interventions}
\label{intervention refusal}
In addition to \textit{guarding information}, seniors rejected their children's interventions in various ways. Psychologically, they expressed \textit{distrust} by believing scammers over their own family. Behaviorally, they \textit{grilt-tripped} younger family members using the concept of filial piety. Cognitively, there was also \textit{a conflict between seniors' autonomy and younger family members' desire to control}, as seniors sought to maintain independence and make their own decisions.

\textbf{\textit{Distrust.}} Seniors often firmly believed the words of scammers while distrusting their own family members and even official institutions such as hospitals and the police. For instance, one individual shared, 

\begin{quote}
\textit{``My mom spent 50,000 on a mattress advertised by a known scam reported in the news. No matter how much we tried to convince her, she chose to trust the scam over her own children. She didn't even consider a visit to the hospital, insisting on trusting these deceivers over established science.''} 
\end{quote}

Additionally, we found that elderly people would protect scammers in front of the police. For example, a daughter was extremely helpless when her mother lied to the police, 

\begin{quote}
\textit{``After confronting her, she still wanted to invest more. I asked the police to intervene, but she told them she hadn't made any transfers, still firmly believing in the scam. The police told me they couldn't help if she didn't admit to being a victim.''} 
\end{quote}

When the seniors themselves were not aware of the scam, seeking redress became an even more difficult task. Moreover, seniors might also suspect that their family members tried to seize their assets, leading to significant frustration and disappointment among younger family members who feel disinclined to intervene further. A younger person said,
\begin{quote}
    \textit{``All my grandfather's money went to a high-risk company. When I asked him to reclaim it, he thought I was trying to take his money. He'd rather trust an obvious scam than his own granddaughter. I've given up now, and he might as well end up begging once he realizes he's been scammed.''}
\end{quote} 

Seniors' distrust could originate from real concerns, as financial exploitation by family members is a common type of fraud encountered by the elderly \cite{deliema2018elder}. Furthermore, seniors would demean their family members and doubt their capabilities, as one individual shared, \textit{``My mom thinks I have no right to criticize her because she believes I wouldn't know any better either. Our conversation tonight ended unpleasantly, and it really made me cry.''} This behavior not only facilitated the scam but also severely damaged family relationships, a point we elaborated in \S\ref{difficult rights protection}.

%filing a report required the elderly person to be aware that they were scammed and filed the report as the victim, otherwise it would not be accepted by the police.
%\textbf{\textit{Refusing anti-fraud information.}} In Section \ref{finding: RQ2}, we demonstrated how younger individuals conveyed anti-fraud information to the elderly, whether through persuasive or educational behaviors, such as sharing similar fraud cases to raise fraud awareness. However, some seniors were resistant or distrustful of this information. For example, one young person mentioned, \textit{``No matter how much I say or even send them videos, the older folks won't watch them. They even dismiss videos made by the police. It's frustrating.''} Moreover, seniors often leveraged their parental authority to degrade the younger generation, e.g., \textit{``Even if parents watch it, they won't believe what you say. They'll just tell you that you don't understand!''}

\textbf{\textit{Guilt trip.}}
To ``assist'' scammers in deceiving themselves, seniors would impose moral pressure on their children, frequently invoking the concept of \textit{filial piety}, which is profound in Chinese culture \cite{ikels2004filial}. Being accused of unfilial behavior carried a significant social stigma and moral burden in China, as filial piety was not merely a personal virtue but a pivotal societal expectation \cite{o2015chinese}. This cultural value could strongly influence younger family members, who might feel compelled to comply with their elders' wishes. For instance, one child expressed their dilemma,
\begin{quote}
    \textit{``It's incredibly real. Whenever you try to convince them with logic and facts, they hit you with accusations of being unfilial or remind you of all the money they spent raising you to manipulate you with guilt. It's truly infuriating.''}
\end{quote}

Beyond such moral manipulation, seniors also tried to evoke their children's empathy, as one individual shared, \textit{``My dad says that buying those fake ceramics is his heartfelt wish in his old age, and he enjoys it using his own money, should I really stop him? He spends thousands from his salary every month anyway.''} This behavior undoubtedly placed children in a difficult position, thereby softening their stance against intervening in scams. 

Moreover, children would compromise to maintain family harmony and ensure their parents' happiness, despite their better judgment, as in the following example:

\begin{quote}
    \textit{``My mom told me she wanted to buy something today, so I did some research and found there's no scientific basis for it. But to avoid any arguments, I went along with her decision. After all, she's unhappy if she doesn't buy it. I just told her not to use it every day and treated it like an occasional health supplement.''}
\end{quote}

\textbf{\textit{Conflicts between seniors' autonomy and younger family members' control.}} Some elderly individuals had a strong sense of ownership and autonomy over their own affairs. When younger family members provided supportive behaviors, the elderly might feel that their independence was being challenged. For example, a young person said that her dad did not allow her to delete anything from his devices: 
    \textit{``It's already great that your parents let you help uninstall things. My dad's attitude is very hostile, he doesn't permit me to delete anything.''}

Furthermore, this sense of autonomy also drove the elderly to refuse younger people's attempts at fraud intervention, using phrases such as \textit{``It's my money, you don't get to manage it. I'm willing to be scammed.''} When seniors exhibited strong independence, younger individuals might choose to reduce their control to avoid potential arguments. As one younger individual shared, \textit{``She's fiercer than anyone. `Am I spending your money? Why do you care!' So, I just don't dare to speak up anymore.''}

\subsubsection{Dark Patterns in Online Payment Systems}
\label{tension}

Although some spending by seniors appeared minor and legally permissible, specific features in online payment systems could facilitate dark patterns. Many mechanisms made payments especially easy, such as password-free payments and ``buy now, pay later'' options. One young person mentioned, 

% \textit{\textbf{Overly convenient payment and dark patterns.}} 

\begin{quote}
\textit{``Many elderly people are induced to enable password-free payments. Orders under 100 yuan don't even show a payment page, and it just places the order directly. Unscrupulous merchants deliberately price items at 99 yuan to skirt the rules.''} 
\end{quote}

Moreover, important information was often minimized or hidden in a way that made it difficult to notice, as one young individual commented, \textit{``The option to enable password-free payments is in fine print, which the elderly can hardly see.''} Platforms also facilitated seamless transitions from one app to the other. One younger person described the experience, 

\begin{quote}
\textit{``Typically, seniors are lured by short videos on platforms like Douyin. It gets cut off at the key moment, and then it jumps to the WeChat payment page. Although watching is free, the rest of the content requires a paid subscription. Elderly people end up paying without realizing it.''} 
\end{quote}

For seniors with low digital literacy, the issue was more severe. This situation raised the question of whether these overly convenient payment methods were genuinely user-friendly or cleverly designed traps. Indeed, young individuals shared their observation that some businesses were exploiting the low digital literacy of the elderly. One person commented,
\begin{quote}
\textit{``It's infuriating to see posts about merchants heartlessly exploiting the elderly, who are unfamiliar with online shopping. They display prices like 9.9 yuan, but the payment page shows 99 yuan. Many don't even realize how much they've spent, and if they do find out, they don't know how to seek redress.''}
\end{quote}

In addition, closing ads could be a challenge with invisible or fake close buttons designed to force accidental purchases. One individual described such features, \textit{``As soon as you open the page, the first thing that appears is an ad with an almost invisible close button in the top right corner, which many older people can't see and end up inadvertently making payments.''} 
%Even the younger generation was not immune to such traps, e.g., \textit{``the order mechanism on Pinduoduo is really tricky. It's too easy to accidentally place an order. I've done it many times myself.''}

% \textit{\textbf{Deceptive inducements}}, especially using red packets. 
%which hold special significance in Chinese culture and ,
Red packets, traditionally given as monetary gifts during special occasions, have evolved into online formats like WeChat Red Packets \cite{wu2017money}. However, these digital adaptations have also been commonly abused as a fraud strategy. These cultural tokens are manipulated through various digital forms, such as \textit{fake and click-baiting red packets} that redirect the user to payment pages, causing older individuals to mistake fraudulent schemes for legitimate online transactions. For example, one young person recounted, \textit{``My mom was just watching a video and suddenly paid for a membership through a red packet link. Older people think it's like the red packets we send on WeChat, so they just click and end up paying unexpectedly.''} 
%Beyond the overt enticement of red packets, there were more subtle forms of manipulation through algorithm-driven recommendations that subtly marketed products to seniors. These algorithms tailored content to the interests of the elderly, making it difficult for them to resist. For instance, one person shared, \textit{``My mom searched for foot pain remedies on Douyin, and now it occasionally recommends ointments to her. They're not cheap, but whatever the videos recommend, she believes.''}

% From the perspective of seniors, the issue of low digital literacy exacerbated the ease with which these scams occurred.  
%Many elderly users were confused and misled into enabling settings that led to unintended charges. 

% , coupled with these convenient payment methods, to deceitfully extract money.

\subsubsection{Limited Consumer Protections}
\label{difficult rights protection}
\S\ref{Loss Recovery} showed how some younger individuals took on the responsibility of recovering losses for the elderly. In this process, they encountered multiple challenges in helping the seniors protect their rights.

\textit{\textbf{Lack of knowledge of reporting channels.}} Many young people did not know how to seek redress for unconscious consumption. A common trend observed in our dataset was young people sought help through social media, either by posting for advice or commenting on similar cases for guidance. For instance,
\begin{quote}
    \textit{\textbf{Post (Help seeking):} ``My mom is not very good at using a mobile phone. I just discovered that there were inexplicable deductions last month. I can't find the public account or anything that the deductions were made to, and the records show no information \dots I checked her identity information and there was no theft or unauthorized subscriptions, so I don't know if I can get this money back.''}
\begin{itemize}
    \item \textit{\textbf{Comment1 (Solution):} ``I just went through almost the same thing. Just file complaints with each service; most will refund directly, some might ask you to contact customer service via WeChat and provide order numbers.''}
    \item \textit{\textbf{Comment2 (Help seeking):}``Just discovered my elderly family member also had mysterious deductions by [merchant name] while watching a series. No idea how to get a refund, what do I do?''}

    \item \textit{\textbf{Comment3 (Update from post author):} ``Thanks for the help, everyone. I've filed complaints on all transactions via WeChat Pay, waiting for further updates.''} 
\end{itemize}

    %\textit{\textbf{Comment 2: (Solution):}``You can file a complaint about the order issue on the payment platform (WeChat, Alipay) by selecting 'Order Complaint'.''}
\end{quote}

Whether in posts or comments, many young people did not know the specific steps to seek redress. Social media partly filled the knowledge gap as it enabled younger family members to collectively brainstorm and find appropriate redress channels, as well as share progress and solutions.

\textit{\textbf{Difficulties in seeking redress.}}
Even when young people knew possible avenues to lodge complaints, they often faced hurdles such as unresponsive customer service or false contact details. One young person vented, 

\begin{quote}
\textit{``Aside from recovering less than a quarter of the amount yesterday, other complaints are either ignored, or they provide fake numbers and fake WeChat customer service accounts that can't be searched, or they just say the service has been used and tell off the complainers.''.}
\end{quote}

\subsubsection{Enormous Mental and Financial Stress}
\label{mental stress and family strain}
The supportive behaviors described in \S\ref{finding: RQ2} suggested that such assistance required young people to invest significant time and effort. These processes created a dual burden of external and internal pressures for the younger generation. 

External pressures include shouldering the family's financial losses and preventing the elderly from being defrauded again. The financial burden alone could be crushing, as one young person complained, 

\begin{quote}
\textit{``The money is in their hands, and no matter what I say, they won't listen. In the end, I have to clean up the mess. I've already helped them pay back over 1 million yuan. What else can I do? From now on, I won't take responsibility for them anymore. I can't keep dealing with this.''}
\end{quote}

Internally, strained family relationships were another major source of significant stress. Many younger family members disclosed feelings of helplessness and disappointment when dealing with this issue. One young person even described these challenges as torturous,
\begin{quote}
    \textit{``I've been tormented for years. I felt utterly helpless and disappointed. Most of our fights have revolved around these issues. My mom is stubborn. She won't listen to me, acts irrationally towards scammers with complete trust, and becomes impatient and aggressive when discussing these matters with me. She refuses to reveal details or admit she's been scammed and even lashes out at me for calling out the scammers.''}
\end{quote}
In severe cases, the elderly might threaten to cut off relationships, as one individual cried for help, \textit{``What do I do? My mom wants to cut off our mother-daughter relationship!''} The strained family relationships had a profound psychological blow to the young generation. Moreover, young people faced a moral dilemma between adhering to social norms such as filial piety and being firm about protecting seniors from fraud. This internal conflict could also significantly contribute to younger family members' mental stress.
% , as noted in \S\ref{intervention refusal}.
The stress further led to insomnia (\textit{``I can't sleep. I'm having trouble sleeping.''}) or a sense of exhaustion and resignation (\textit{``I'm tired of it, I won't let them make me pay the money anymore.''})

\section{Discussion}
In summary, our work provides insights into senior-targeted online fraud, younger family members' support and challenges based on a comprehensive analysis of posts and comments on RedNote. In this section, we situate our findings in prior literature, reflect on the whole support ecosystem against senior-targeted online fraud, and propose implications and recommendations for safeguarding seniors from online fraud.

%discuss our research's limitations, 
\subsection{\modi{Situating Findings within the Literature}}
\label{Comparisons with Prior Work}
%Revisit Senior-targeted Online Fraud from the Lens of Family Members in China
% Extensive research has looked into the significant role of family members in  digitally disadvantaged older adults.

%\dy{Previous studies have highlighted that older adults are particularly susceptible to fraud, with many falling prey to various schemes specifically targeting their vulnerabilities \cite{url25, oliveira2017dissecting,grimes2007email}. Our work builds on this understanding by offering a taxonomy of these vulnerabilities, identifying that multiple types of fraud often exploit similar underlying vulnerabilities among older individuals.}

%~\cite{url25, oliveira2017dissecting,grimes2007email}
\subsubsection{Unique help enabled by younger family members.} \modi{Our findings revealed that younger family members could help identify diverse fraud types faced by older adults.} Aligning with prior work, we identify common types of scams targeting seniors, such as \textit{investment fraud} \cite{url25,button2009better}, \textit{romance fraud} \cite{buil2022meeting,cross2016they} and \textit{fake payment} \cite{he2023have} \modi{shown in Table \ref{table1}}. Beyond these similarities, we extend prior work by revealing fraud types that older adults struggle to recognize, such as \textit{unnoticed charges} and \textit{false advertising of collectibles} in \S\ref{finding: RQ1}. \modi{Moreover, younger family members support recognizing various fraud details. Previous work showed older adults often did not fully understand how their behaviors might expose them to risks \cite{oliveira2017dissecting, nicholson2019if}. We found that younger family members unveiled the specific channels and strategies used by fraudsters, such as \textit{exploitation of social relationships} (e.g., tricking seniors via building connections on WeChat in \S\ref{finding: RQ1}) and \textit{dark patterns} in online payment systems (e.g., \textit{fake red packets} mentioned in \S\ref{tension}). These complicated and manipulative mechanisms can be challenging for the elderly to detect on their own.} This provides empirical evidence supporting previous work that many victims may not realize they have been scammed for a long time, and it is often family members who see through the deception \cite{sivagumaran2023challenges, gloag2019protected}.
%Much previous research has looked into how family members support older adults' technology use and finances, such as family's support in financial technology onboarding~\cite{tang2022never} and family's account sharing for digital inclusion~\cite{he2023have}. Our study further extends this line of work to family's safeguarding of older adults' online financial security.
%Specifically, in \S\ref{finding: RQ2}, we reveal the multi-faceted and dynamic role younger family members play in fostering seniors' resilience to fraud, ranging from pre-fraud prevention and identification to post-fraud loss recovery and education. Moreover, we identify five sorts of challenges faced by younger family members.
%These findings might help inform future anti-fraud strategies, potentially guiding police investigations into senior-targeted fraud, providing possible online moderation practices, supporting new education materials for fraud awareness campaigns and shaping policies on fraud prevention by governments.
\modi{To this end, we suggest future research take a holistic view to examine older adults' online security, recognizing the sociocultural contexts including the support and influence of family models.}

%Integrating the perspective of family members could help to identify more categories of senior-targeted fraud and provide greater detail on how seniors are scammed. 
%informing the design of fraud prevention tools,
%Moreover, younger family members discovered \textit{dark patterns in online payment systems}, such as \textit{fake red packets} and \textit{minimized or hidden payment-related information} in \S\ref{tension}.

%we also found that multiple types of fraud often exploit similar underlying vulnerabilities among older individuals

%Repercussions on young family members themselves in a collectivist culture. 
\subsubsection{\modi{Strong family bonds in Chinese sociocultural contexts}}
\textit{Filial piety} \cite{ikels2004filial} is an important concept in Chinese culture, which sets up the expectations for children to care, respect, and show obedience to older family members. Filial piety contributes to closer ties between family members in Chinese families compared to those in Western countries~\cite{tang2022never}. \modi{Tang et al. found that filial piety helped seniors feel comfortable seeking help from younger family members in technology support, viewing their protection as beneficial rather than disempowering~\cite{tang2022never}. However, our findings revealed filial piety functioned as a double-edged sword in online fraud. While it served as a foundation for family support in anti-fraud efforts, seniors leveraged filial piety to compel younger family members to accommodate scammers' requests in \ref{intervention refusal}, causing enormous pressure on younger family members. Moreover, when seniors trusted scammers over their families, it could disappoint younger members and halt their support in \S\ref{intervention refusal}. Given this, we suggest future anti-fraud educational campaigns fostering a balanced understanding of filial piety to encourage mutual respect and communication within families and raise awareness about its potential misuse in facilitating online fraud.}
%\S\ref{mental stress and family strain} reveals that children experience mental stress and tensions in family relationships during the support process.
%We call for future research to look into how to improve younger family members' well-being and alleviate their stress.
% Part of the stress could come from 

%and we discussed some potential approaches in \S\ref{Implications and Recommendations}.

%\modi{A family-oriented model to counter senior-targeted fraud brings additional challenges, such as older adults' \textit{guard information} for various reasons like fear of blame and perceived superiority, and \textit{refusal of help and interventions} in \S\ref{information segregation} and \S\ref{intervention refusal}. Moreover, 
%Our study adds more nuances to prior work by showing that filial piety can work as a double-edged sword in a family's technology use. 
%%protect and care for their elders while emphasizing respect and obedience to parents, coupled with the closer familial ties typical in Chinese families compared to those in Western countries \cite{tang2022never}, may significantly contribute to these pressures. 
%These findings echo prior work which argues that family support in technology use in China may bring asymmetry due to hierarchical parent-child relationship~\cite{tang2022never,he2023have}. 

\subsubsection{Tension between financial security and autonomy.} Financial autonomy, beyond access, has become a crucial factor in financial inclusion~\cite{muralidhar2019rethinking}. However, our work cautions against the possibility of reduced financial autonomy when older adults are exposed to online fraud and family members take strict paternalistic approaches such as \textit{online payment restriction} and \textit{monitoring} in \S\ref{prevention behavior}. %Similar to Murthy et al.~\cite{murthy2021individually}, we found that paternalism in protective behaviors could negatively impact older adults' agency. 
%Similar to Murthy et al.'s study on family support in general online security and privacy, where family members sometimes adopt controlling approaches to enhance older adults' security, our findings also highlight that younger family members often use controlling methods to prevent older adults from falling victim to fraud in \S\ref{prevention}. However, unlike Murthy et al.'s findings that older adults generally prioritize collective security over individual privacy and are open to family members managing their devices \cite{murthy2021individually}, our study reveals that in the context of online fraud, older adults often display behaviors such as guarding information and refusing help and intervention (e.g., distrusting younger family members) as discussed in \S\ref{information segregation} and \S\ref{intervention refusal}.
%prioritize collective security over individual privacy and 
%(e.g., distrusting younger family members) 
\modi{Different from Murthy et al.'s study on family support in general online security and privacy, which found that older adults are generally open to family members managing their devices and accounts \cite{murthy2021individually}, our findings reveal that in the context of online fraud, older adults often exhibit behaviors such as \textit{guarding information} and \textit{refusing help and intervention} in \S\ref{information segregation} and \S\ref{intervention refusal}. We encourage future research to propose approaches to balance financial autonomy with financial security when designing anti-fraud intervention systems.}

\subsection{\modi{Theoretical Insights into the Support Ecosystem against Senior-Targeted Online Fraud}}
\label{5.1}

\begin{figure}[htbp]
	\centering
	{\includegraphics[width=1\columnwidth]{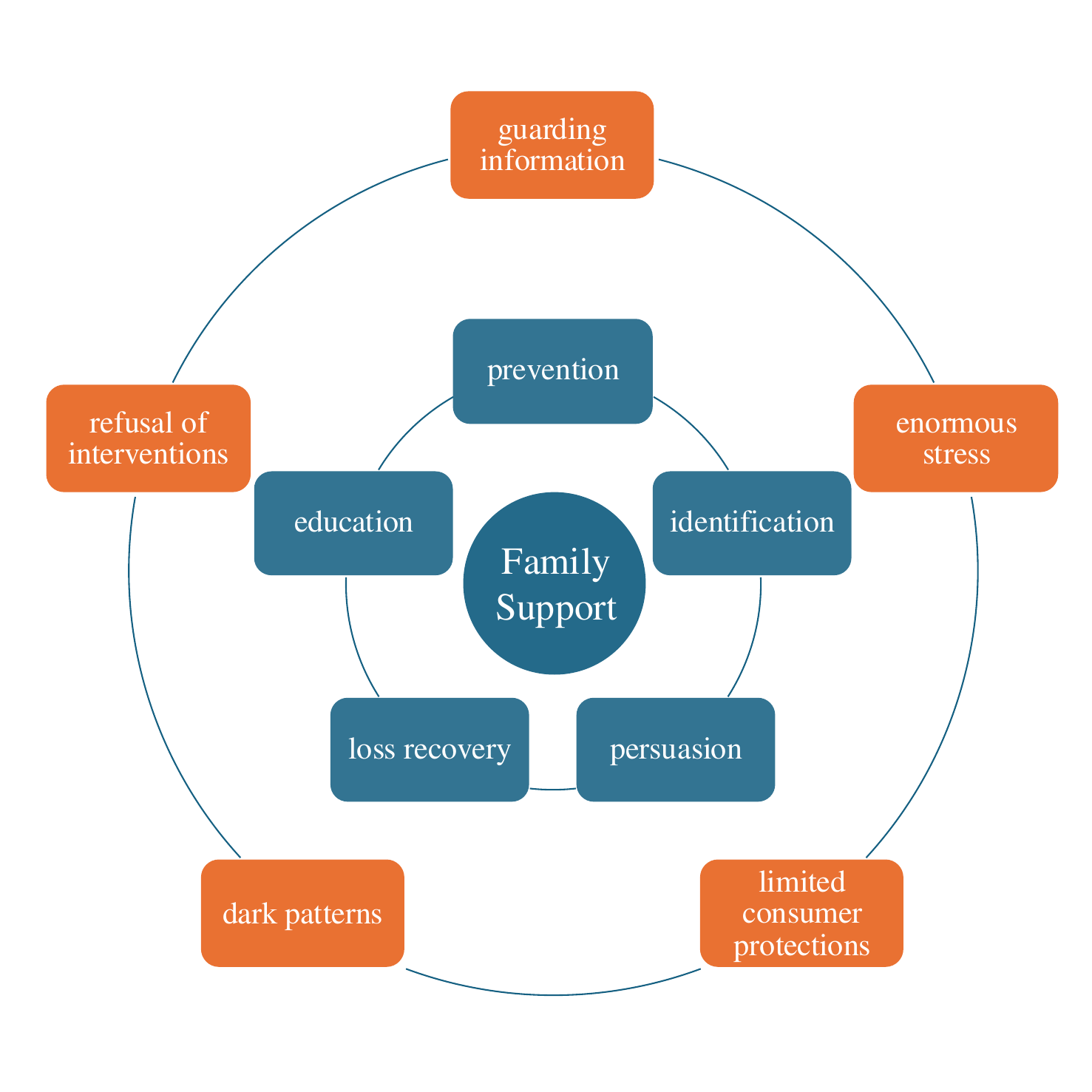}}\\
  \caption{\modi{The conceptual framework of family support against senior-targeted fraud. The supportive behaviors are highlighted with the blue color and the challenges are showcased with the orange color.}}
  \Description{Figure 3: This figure shows the conceptual framework of the family support ecosystem in safeguarding seniors against online fraud. Family members play multiple supportive roles including prevention, identification, persuasion, loss recovery and education. Beyond family members, there are also other stakeholders such as official departments, the local community and social media, although they may not directly connect with seniors.}
  \label{FIG: framework}
\end{figure}

Combating senior-targeted fraud involves efforts beyond the victim \cite{deliema2023correlates,parti2023if,nicholson2019if,berry2013effect}. The family systems theory \cite{bowen1993family} suggests that when addressing a crisis, individuals cannot be fully understood in isolation, but rather as components of a family unit. In the context of online fraud, our findings provided empirical evidence, demonstrating that family members indeed played multiple roles, from somewhat controlling behaviors to prevent the fraud to identifying and persuading seniors about the fraud's existence, providing both financial and emotional support in the recovery process and educating the elderly about fraud (\S\ref{finding: RQ2}). \modi{Our empirical findings in \S\ref{finding: RQ2} and \S\ref{findings: RQ3} inform a conceptual framework of family support in safeguarding seniors against online fraud, as outlined in Figure \ref{FIG: framework}. Except for these basic elements, family support also displays three important characteristics: }
% In \S\ref{finding: RQ2}, we found they helped \textit{prevent} fraud, \textit{identified} scams, \textit{persuaded} seniors of the fraudulent nature of scams, assisted in both financial and emotional \textit{loss recovery}, and actively participated in \textit{educating} the elderly about fraud. 
 %and is intricately woven into seniors' ongoing resistance against scam activities. 
%When addressing a crisis, it is crucial to consider the family as a whole rather than focusing solely on only one family member. 

% From a perspective of safeguarding seniors against online fraud, 

% , showing how families counter senior-targeted fraud as an organic whole. 
% 
\modi{(1) \textit{long-term and evolving}. Our findings reveal that family support for seniors in online fraud scenarios is prolonged and evolving. For example, the three-phase identification process in \S\ref{identification}, the years-long persuasion efforts described in \S\ref{mental stress and family strain}, and daily emotional care shown in \S\ref{Loss Recovery} highlighted the sustained nature of such support. Seniors may also fall victim to fraud multiple times, requiring repeated family support as described in \S\ref{mental stress and family strain}. Furthermore, younger family members adapt and refine their support behaviors over time. For instance, they may seek new redress solutions by leveraging advice from social media users in \S\ref{difficult rights protection} or update their educational materials in response to emerging fraud cases in \S\ref{Education: Active Involvement in Fraud Education}. Given this long-term and evolving feature, we call for future work to explore the longitudinal dynamics of family support, such as the evolution of supportive behaviors.}
%Given the long-term and evolving features, it is crucial to explore ways to better support younger family members throughout this process. Additionally, future research could investigate how to facilitate the sharing of accumulated knowledge and experiences among families} 
 
%Many younger family members reported being burdened for extended periods by their seniors' victimization, experiencing relentless persuasion efforts or repeatedly helping repay debts.

%In \S\ref{Implications and Recommendations}, we propose possible ideas for a dedicated senior-targeted online forum to address these needs.}

%However, this process may also halt due to frustrations from \textit{seniors' refusal of help} discussed in \S\ref{intervention refusal}. 
%The support requires collaborative efforts from 
\modi{(2) \textit{multiple stakeholders}. Younger family members often serve as mediators, bridging the family with various external stakeholders involved in anti-fraud efforts. For example, younger family members may turn to the police for assistance in persuading older adults, as shown in \S\ref{persuasion}. \S\ref{difficult rights protection} highlighted the difficulties in seeking redress through customer service, and social media provided a platform for younger family members to seek help and discuss possible solutions. These examples depicted younger family members as the primary mediators in obtaining support from other stakeholders. Future work should explore strategies to better equip younger family members with resources and tools to facilitate their mediation role effectively while also examining how other stakeholders can collaborate more seamlessly within this support ecosystem.}

%Future research could explore ways to enable more direct interactions between older adults and these stakeholders. This could not only reduce the burden on younger family members but also foster older adults' trust and awareness of anti-fraud efforts across various sectors.
%challenges posed by limited consumer protections, such as 
%such as police, customer service representatives, or social media
% to enhance their overall resilience to fraud
%Family members often work  by orchestrating interventions with the police to persuade parents or sharing new online fraud cases learned from social media shown in \S\ref{persuasion}

\modi{(3) \textit{latency in support}. There may be a mismatch in timing between the support provided and the occurrence of the fraud. For instance, by the time younger family members recognize the fraud, financial losses may have already occurred, as described in \S\ref{identification}, necessitating post-fraud recovery efforts. This latency is further exacerbated by seniors' \textit{guarding of information} shown in \S\ref{information segregation}. Therefore, future work could explore ways to conduct timely interventions from family members, reducing losses and alleviating the subsequent burden on younger family members.}

%We propose a platform intervention as a potential solution to enable more timely and effective support in \S\ref{Implications and Recommendations}.
% in support 
%younger family members may sense something is wrong during persuasion attempts, yet seniors may continue to fall deeper into the scam. Similarly, 
%In some cases, seniors may have already been defrauded by the time family members become aware, 

Our conceptual framework provides a novel perspective of family support on senior-targeted fraud. We encourage future work to empirically validate and expand upon this framework.
%However, this family support ecosystem still faces challenges,
%

%
%Previous research on anti-fraud frameworks typically focused on business organizations, emphasizing strategies to counter corporate fraud may be committed by corporate officers or employees \cite{albrecht2006fraud,todorovic2020anti}, such as prevention, detection, knowledge and action \cite{mangala2017auditors}. 
%(2) \textbf{\textit{Potential unequal involvement across phases.}} Although family support includes various supportive behaviors, the level of involvement can vary. For instance, there might be extensive effort in \textit{prevention} but less in \textit{education}, or minimal \textit{identification} but substantial \textit{loss recovery}.

%and we advocate for broader stakeholder engagement
%Moreover, our conceptual framework is informed by a social media dataset that consists of discrete posts and comments, so we cannot capture complete narratives of the support cycle. and explore how various supportive behaviors evolve over time and as new fraud schemes emerge

% \subsection{Revisit Senior-targeted Online Fraud from the Lens of Family Members in China}

\subsection{\modi{Practical Implications and Recommendations}}
\label{Implications and Recommendations}
\subsubsection{Platform interventions}
\modi{In \S\ref{Comparisons with Prior Work}, we discussed the necessity of balancing online autonomy with financial security.} Moreover, our findings revealed sometimes older adults actively sought assistance from younger family members, such as asking their children to check their accounts, as shown in \S\ref{identification}. \modi{Given this, designers
can develop mechanisms that empower seniors to easily request help when needed, satisfying seniors’ need for financial autonomy.} One potential approach is to integrate a built-in tool on seniors’ mobile phones that facilitates seeking advice from family members. \modi{For instance, older adults encountering financial security concerns could click the tool to automatically capture screenshots, highlight possible problematic areas, generate concise descriptions of the potential issue, and revise under the tool's guidance if the content is inaccurate by using advanced technologies (e.g., generative AI). The summarized information could then be forwarded to a chat interface (e.g., WeChat) for family members to review and respond. Alternatively, with the senior’s consent, younger family members could remotely access the device to take necessary actions or demonstrate solutions through the tool.} Such a built-in tool has the potential to enhance online safety for seniors while also reducing tensions between seniors and younger family members. 

%Visual aids (e.g., highlighted pens) and audio explanations could guide the older adult through the process. This mode could be especially helpful for older adults who do not live with their family members. 
%However, \S\ref{information segregation} suggested older adults would guard their information from younger family members, 

%Nevertheless, a purely younger family member-initiated approach might also be problematic, as it could evolve into excessive control over older adults, such as \textit{preventing online activities} entirely in \S\ref{prevention behavior}, significantly limiting their autonomy.
\modi{However, our findings also revealed that sometimes older adults were reluctant to share their online activities and guarded their information in \S\ref{information segregation}, which indicates that an older adult-initiated approach may not always work effectively. Additionally, \S\ref{5.1} highlighted \textit{the latency in support}, emphasizing the importance of timely intervention from younger family members for better protection. To this end, designers could consider developing a system that automatically alerts younger family members to intervene promptly under specific circumstances, ensuring timely support while maintaining a balance between autonomy and security. One feasible solution could involve a collaborative approach where older adults and their younger family members set a predefined monetary threshold for interventions. When a payment exceeds this threshold, the system could trigger different actions depending on different payment stages. For instance, before payments, it could actively prompt older adults, asking whether they wish to discuss the payment with a younger family member, summarize payment details (e.g., product descriptions, payment links) and share them through a chat interface for discussion. During transactions, the tool could require explicit verification from younger family members, temporarily blocking the transaction until approval is granted. After payments, it could send notifications of transactions beyond the threshold to younger family members, enabling them to follow up as necessary. For specific designs, we call for future research to focus on co-designing such support models through workshops or participatory design, involving older adults, their families, and other stakeholders like caregivers and communities to identify older adults’ needs for flexible support models while ensuring online autonomy and financial security. Generalization of design implications to other countries may require consideration of family dynamics within the local sociocultural context.}

\modi{In addition to the intergenerational support model, there are other design implications worth considering. For instance, \S\ref{tension} highlighted the presence of \textit{dark patterns}, such as \textit{overly convenient password-free payment options, fake close buttons}, etc. Relevant policies should explicitly address these issues to promote accountability}. Additionally, we observed that \textit{red packets} were exploited in various fraud types, indicating their significant appeal to seniors (Table \ref{table1}). A possible design solution is a warning message in interfaces involving red packets to alert seniors. We suggest researchers and designers systematically evaluate how fraud-checking nudges in online payments could prompt seniors to think carefully before proceeding.

%For example, virtual accounts generated by AI could interact with older adults, mimicking fraudsters to test their responses and provide educational feedback to raise awareness. 
\subsubsection{Education for older adults}
%Future research could examine the feasibility of this approach, but it also has potential risks, such as whether prolonged interactions might lead older adults to over-engage with or believe the false narratives created by these simulations.
%we suggest that educational materials should go beyond outdated and overused slogans like ``There is no such thing as a free lunch''. 

\modi{Drawing on our findings about older adults' \textit{guarded behavior} and \textit{refusal of intervention} in \S\ref{findings: RQ3}, educational materials should highlight the critical role of family members in anti-fraud. In the Chinese context, where filial piety and the authoritative or authoritarian position of seniors are ingrained \cite{chuang2018parenting}, educational efforts should challenge superiority-based perspectives that undermine younger family members' roles in recognizing and preventing online fraud. For example, new anti-fraud guidelines could explicitly showcase that \textit{distrusting} family members and \textit{guarding information} from younger family members shown in \S\ref{information segregation} and \S\ref{intervention refusal} are the common brainwashing tactics used by scammers. Prior research also noted that US older adults were often unwilling to disclose fraud experiences to family members due to concerns
about losing family status, trust issues and others \cite{parti2023if}, indicating that this implication might hold value for other contexts as well.}
%Furthermore, older adults should not use \textit{guilt trip} in \S\ref{intervention refusal} to pressure younger family members into compromising or refraining from intervening in fraud-related situations. By breaking free from these guilt dynamics and fostering open, rational discussions, families can better collaborate to prevent fraud.}

%should be explicitly highlighted to help both younger family members and older adults recognize that seniors may, consciously or unconsciously, use this strategy 

%in HCI field has already simulated phishing emails to investigate users’ susceptibility to phishing, without disclosing the research goal to participants until the experiment end \cite{lin2019susceptibility}.

\modi{Additionally, an interesting phenomenon was observed in \S\ref{Education: Active Involvement in Fraud Education} that younger family members simulated fraudsters to raise seniors’ awareness of potential fraud. Existing research has also explored fraud simulation as a method to directly capture the behavioral response to attempted fraud ``in the wild'' \cite{lin2019susceptibility}, providing an immersive and realistic experience. Inspired by this, future work could explore simulating real-life fraud scenarios through customized education tools tailored to various fraud types, channels, and strategies based on \S\ref{finding: RQ1}, enhancing seniors' comprehensive fraud prevention literacy. Future designers could consider potential simulation formats, such as gamified fraud simulations or AI-based real-life simulations. For example, \textit{exploitation of social relationships}, a common fraud tactic mentioned in \S\ref{finding: RQ1}, often involves interactive relationship-building, which could be simulated using AI to educate seniors. However, designers must consider ethical issues. For instance, explicitly labeling it as an education tool might reduce its effectiveness, as older adults could develop preemptive awareness, while not disclosing this could create artificial stress or be perceived as deceptive, even if deception has been used in previous work \cite{lin2019susceptibility}. The use of AI introduces another crucial question regarding whether and how to disclose its involvement. Future research should also pay attention to whether prolonged interactions might lead older adults to over-engage with or believe the false narratives created by these simulations.}

%with the assistance of AI .

%Our work uncovered various specific online fraud types and fraud details in \S\ref{finding: RQ1} that could inform .

% Additionally, the implementation format, whether through gamification, real-life simulation or others, requires careful consideration and investigation.

%In \S\ref{Education: Active Involvement in Fraud Education}, we identified that younger family members educated seniors about fraud from \textit{cognitive}, \textit{psychological}, and \textit{operational} perspectives. However, seniors' fraud education should extend beyond family members to involve more stakeholders, such as police and local communities. While our conceptual framework includes different stakeholders, many act indirectly through family members. We call for these stakeholders to take more direct and collaborative actions. For instance, police could share recent fraud cases with local communities, and local communities could host offline workshops and establish social app-based groups to promote fraud cases to seniors \cite{kropczynski2021towards}.

%One of them is the lack of knowledge about reporting channels in \S\ref{difficult rights protection}.
\subsubsection{Support needed for younger family members}

\modi{In \S\ref{findings: RQ3}, we found that the supportive process involves numerous challenges faced by younger family members.} Our findings in \S\ref{difficult rights protection} showed younger family members could discover potential solutions on social media. \modi{Additionally, social media allows them to gather new fraud cases, which could be used to educate seniors about emerging threats mentioned in \S\ref{Education: Active Involvement in Fraud Education}. Therefore, facilitating the use of social media to seek help and discuss online fraud-related solutions might benefit more families. To this end, future designers could consider creating a dedicated discussion space on social media specifically for senior-targeted anti-fraud, allowing information to be more centralized and accessible. Platforms may consider how to prevent discussions from being exploited by scammers, such as implementing robust authentication mechanisms, to ensure a secure and trustworthy environment.}

\modi{Furthermore, we call for the participation of more stakeholders in the discussion space on social media. As highlighted in \S\ref{Education: Active Involvement in Fraud Education}, one notable challenge younger family members face is strained family relationships, particularly when efforts to persuade seniors result in conflict. A compelling example from \S\ref{persuasion} involved a younger family member ``@'' the older adult in a discussion thread, prompting another user to assist in persuasion. To alleviate the emotional stress on younger family members, inviting more stakeholders, such as police representatives, community members or psychologists, to join the conversations might make it easier to convince seniors and probably foster harmonious family relationships. Additionally, the participation of policymakers and law enforcement officials could also be beneficial.} For instance, as illustrated in \S\ref{difficult rights protection}, family members faced significant challenges when seeking external help—e.g., filing a report was not accepted by the police if the elderly did not acknowledge being scammed. \modi{Policymakers could use related insights from social media discussions to address practical challenges and develop human-centric strategies. While specific discussions on social media are rooted in the unique cultural and policy context of China, these design implications, such as the dedicated discussion space on social media and the participation of more stakeholders, could be adapted to benefit other regions.}

\subsection{Limitations}
\label{limitations}

%although exploring senior-targeted online fraud through the lens of younger family members provides a novel and meaningful perspective, 

\modi{First, younger family members may selectively highlight roles or actions that are more favorably perceived by their audience, potentially influenced by social desirability bias \cite{fisher1993social}. Additionally, our data may not include all viewpoints, such as those of seniors, local communities, and official departments. We advocate for future research to adopt a multi-stakeholder approach, incorporating diverse perspectives and data sources to better understand younger family members' roles and seniors’ experiences with fraud.} Second, while we utilized social media to gather a diverse range of viewpoints, our sample might not be fully representative of all older adults in China. We call for future work to validate and cross-compare our findings by exploring a larger corpus of data through quantitative or computational methods. Third, while the perspectives of younger family members provide valuable supplementary insights into senior-targeted fraud, it is important to clarify that this approach is not necessarily superior to direct interviews with older adults. Rather, it offers an alternative lens to enrich our understanding with different nuances. Fourth, as our research was conducted on a Chinese platform, some findings might not generalize globally. Fifth, our dataset shares common limitations with all analyses based on social media data. We cannot fully ensure the credibility of each post and user profile. If the posts contain false or exaggerated information or users fill in fake profiles, we might not be able to detect it. \modi{Moreover, due to potential changes in dataset collection stemming from RedNote's contextual similarity-based search mechanism, the repeatability of this research cannot be guaranteed. However, our findings are based on a thematic analysis of the entire dataset rather than conclusions from individual posts, which helps mitigate the credibility issue and ensure the representativeness of the findings.} We hope future work will validate these findings through more empirical methods, such as interviews and surveys. Finally, the demographic characteristics of RedNote's users, predominantly young and female \cite{url13}, could have influenced the findings reported in our study.

\section{Conclusion}
Family support is a significant component in senior-targeted online fraud. This work makes the first attempt to investigate the younger family members' roles in safeguarding seniors from online fraud. To achieve this, we conducted a thematic analysis of 124 posts with 16,872 comments related to family support against senior-targeted fraud on RedNote, one of the largest life-sharing platforms in China. Our results highlight seniors' vulnerabilities in online fraud, younger family members' supportive behaviors across different stages, and challenges in family support. To conclude, we propose a conceptual framework of family support in senior-targeted online fraud and suggest actionable implications for senior's online security within the multi-stakeholder ecosystem.

\bibliographystyle{plain}
\bibliography{sample-base}

% \section{Vulnerabilities and Online Senior-targeted Fraud Disclosed by Younger Family Members}
%\renewcommand{\arraystretch}{1.6}

%%%%%%%%%%%%%%%%%%%%%%%%%%%%%%%%%%%%%%%%%%%%%%%%%%%%%%%%%%%%%%%%%%%%%%%%%%%%%%%%
\end{document}